\documentclass{aa}
\usepackage{graphicx}
\usepackage{txfonts}
\usepackage{natbib}
\usepackage{longtable}
\usepackage{multirow}
\begin{document}

   \title{Thermo-dynamic and chemical properties of the Intra-Cluster Medium}


   \author{Alberto Leccardi
          \inst{1,2}
          \and
          Mariachiara Rossetti\inst{2}
          \and
          Silvano Molendi\inst{2}
          }


   \institute{Universit\`a degli Studi di Milano, Dip. di Fisica, via Celoria 16,
              I-20133 Milano, Italy
         \and
             INAF-IASF Milano, via Bassini 15, I-20133 Milano, Italy\\
             }



 \abstract
   {}
    {We aim to provide constraints on evolutionary scenarios in clusters. One of our main goals is to
    understand whether, as claimed by some, the cool core/non-cool core division is established once and
    for all during the early history of a cluster.
}
    {We employ a sample of $\simeq$ 60 objects to classify clusters according to different properties:
     we characterize cluster cores in terms of their thermo-dynamic and chemical properties and
     clusters as a whole in terms of their dynamical properties.}
    {We find that: I) the vast  majority of merging systems feature high entropy cores (HEC);
     II) objects with lower entropy cores feature  more pronounced metallicity peaks than objects with
     higher entropy cores.
     We identify a small number of medium (MEC) and high (HEC)  entropy core systems which, unlike most
     other such objects, feature a
     large central metallicity.  The majority of these outliers are mergers, i.e. systems far from
     their equilibrium configuration. }
   { We surmise that medium (MEC) and high (HEC) entropy core systems with a large central
     metallicity recently evolved from low entropy core (LEC) clusters
     that have experienced a heating event associated to AGN or merger activity.}

   \titlerunning{Thermo-dynamic and chemical properties of the ICM}
   \authorrunning{A. Leccardi, M. Rossetti \& S. Molendi}
  \maketitle
%

\section{Introduction} \label{sec: intro}
The classification of objects is an important step in the construction of viable physical models. This is
particularly true for disciplines like astrophysics where, due to the impossibility of
performing measurements under  a set of rigorously controlled conditions, evolutionary schemes are
inferred primarily by comparing  properties observed in different objects.
Galaxy clusters are no exception to this rule, as for other astrophysical sources, much of the early work
has concentrated on establishing a taxonomical framework. In the optical band
classification schemes are based on the richness \citep{abell58,zwicky68} and on the morphological
properties which have been found to correlate with the dynamical state of the systems \citep{abell65,abell75a}.
In X-rays, classification attempts are generally focused on core properties
for the rather obvious reason that cores are the regions more easily accessible to observations.
Most workers concentrate on defining indicators discriminating between cool core (hereafter CC)
and non-cool core (hereafter NCC) systems; these indicators are typically based on
estimates of the intensity of the surface brightness peak  \citep{Vikh07}, the temperature \citep[e.g.][]{Sn06}, the cooling time \citep[e.g.][]{baldi07} or the
entropy \citep[e.g.][]{Ca09} of the central regions of clusters.
There have been attempts to derive dynamical properties of cluster ensembles from X-rays,
the best known example is perhaps that of the power-ratios technique \citep{buote95,buote96}.
Comparison of  core with morphological properties,  as identified with the power-ratios technique,
shows that more disturbed systems tend to have less defined cores \citep{buote96,bauer05}.
One should however keep in mind that the
characterization of the degree of relaxation of clusters through X-ray morphology is
limited by projection effects \citep[e.g.][]{jeltema08} and that
the limited amount of information available at large cluster radii, beyond $\simeq 0.2 R_{180}$,
provides a further complication.
There have been some attempts to compare dynamical and core properties on individual objects
\citep[e.g.~A2034,][]{Ke03}, sometimes
using information from different energy bands \citep[e.g.~A1644,][]{Re04}.

Classification schemes based on chemical properties of clusters have enjoyed considerably less
attention. We have known for some time now that CC clusters, formerly known as cooling-flow
clusters, feature significant central abundance excesses \citep{grandi01} and that the amount of iron associated
with these excesses is consistent with being produced from the BCG galaxy invariably found at the center of
CC systems \citep{grandi04}.
Interestingly, there has been no attempt so far to classify clusters by simultaneously making use of thermo-dynamical
and chemical properties.
In this paper we employ a medium size sample, $\simeq$~60 objects, to address the issue of cluster classification from
various angles, more specifically we will:
 1) define an entropy indicator to classify cluster cores with respect to outer regions;
 2) provide a dynamical classification based on radio, optical and X-ray properties;
 3) compare core and dynamical properties;
 4) compare, for the first time, the entropy based classification with a chemical
   classification.
As we shall see, our classification work will allow us to gain considerable insight in how
cluster dynamical, thermo-dynamical and chemical properties relate to each other.
Another interesting result that will emerge from our analysis is the relevance of outliers, i.e. objects
that fall outside of the distributions defined by the majority of systems, in constraining the
evolutionary processes that shape galaxy clusters.

The breakdown of the paper is the following. In
Sects.~\ref{sec: sample} and \ref{sec: analysis} we describe respectively the sample selection and the
data analysis. In Sect.~\ref{sec: entropy} we provide an account of how our entropy and cooling time
indicators have been constructed. In Sect.~\ref{sec: class sch} we describe two classification schemes based
respectively on core and dynamical properties while in Sect.~\ref{sec: entr vs alt} we compare them
with the  entropy based classification scheme defined in Sect.~\ref{sec: entropy}. In Sect.~\ref{sec: Z scheme}
we define a classification scheme based on chemical properties and compare it to the other classification
schemes discussed in the paper. Finally in Sect.~\ref{sec: concl} we summarize our main findings.

Quoted confidence intervals are 68\% for one interesting parameter (i.e. $\Delta\chi^2$~=~1),
unless otherwise stated.
All results assume a $\Lambda$CDM cosmology with $\Omega_\mathrm{m} = 0.3$,
$\Omega_\Lambda = 0.7$, and $H_0$~=~70~km~s$^{-1}$~Mpc$^{-1}$.

\section{The sample} \label{sec: sample}
Starting from the  \emph{XMM-Newton} archive we selected a sample of hot clusters
($kT > 3$~keV). The redshift spans between 0.03 and 0.25 and the galactic latitude
is greater than 20$^\circ$.
Among those clusters satisfying the above selection criteria, we retrieved all observations
performed before March 2005 (when the CCD6 of EPIC MOS1 was switched
off\footnote{http://xmm.esac.esa.int/external/xmm\_news/items/MOS1-CCD6/\\ index.shtml}) and
available by the end of May 2007.
In Table~\ref{tab: sample} we list the observations of all clusters we analyzed and report
cluster physical properties (e.g. redshift and temperature) and observational details
(e.g. total exposure time and filter).
The redshift value (from optical measurements) is taken from the NASA Extragalactic
Database\footnote{http://nedwww.ipac.caltech.edu}; $kT_\mathrm{OUT}$ is derived
from our analysis (see Sect.~\ref{sec: sp ana}).
Observations are performed using THIN1 or MEDIUM filters.
We excluded from the sample observations that are badly affected by soft proton flares, so
that the total (i.e. MOS1+MOS2+pn) exposure time for all observations is at least 20~ks.
We also excluded observations of extremely disturbed clusters for which it was impossible
to define a center; for what concerns double clusters, we analyzed only the brighter of the
two subunits. The total number of objects surviving our selection procedures is 59,
about half of our objects are local, $0.03<z<0.1$, while the other half is located at
intermediate redshifts, $0.1<z<0.25$.

Although not complete, we have reason to believe that our sample  is fairly representative
of the cluster population as a whole. Indeed objects at $z>0.1$ are
extracted by applying a redshift cut at $z=0.25$ from the sample analyzed in \citet{leccardi08a,leccardi08b},
which we showed to be unaffected by substantial biases \citep{leccardi08b}.
Similarly the low redshift half of our subsample
is extracted from the \citet{edge90} flux limited sample by applying neutral cuts such as throwing away
observations that are badly affected by soft proton flares. The only selection introducing
a certain amount of bias is the one against extremely disturbed clusters for which it was impossible
to define a center. We note that only 4 systems were excluded in this manner and
that they will be reintroduced in a subsequent paper \citep{Ross09} where we will
make use of the 2D information provided by X-ray maps.

\section{Data analysis}  \label{sec: analysis}
In this section we provide some information on the data analysis, a more thorough description
may be found in \citet{leccardi08a} and \citet{leccardi08b}. Here we recall some general
issues and highlight a few differences between the analysis performed in this paper
and the one conducted in our previous works.

\subsection{Event file production}  \label{sec: spec prep}
Observation data files (ODF) were retrieved from the \emph{XMM-Newton} archive and
subject to standard processed with the Science Analysis System (SAS) v7.0.
The soft proton cleaning was performed using a double filtering process, first in a hard
(10-12~keV) and then in a soft (2-5~keV) energy range.
The event files were filtered according to \verb|PATTERN| and \verb|FLAG| criteria.
Bright point-like sources were detected, using a procedure based on the SAS task
\verb+edetect_chain+, and excluded from the event file (see \citealt{leccardi08b} for details).

The $R_\mathrm{SB}$ indicator, i.e. the ratio between surface-brightness calculated inside
and outside the field of view (see Eq.~1 in \citealt{leccardi08b}), allowed us to exclude
extremely polluted observations
and to quantify the quiescent soft proton (QSP) component surviving the double filtering process.
Since local clusters fill the whole EPIC field of view, contrary to what we did in \citet{leccardi08b},
we measured $R_\mathrm{SB}$ above 9~keV where EPIC effective areas rapidly
decrease.
We excluded all clusters for which the mean $R_\mathrm{SB}$ for the two MOS is greater than
2.0. The threshold is higher than that chosen in \citet{leccardi08b} because the aim of this work is the
analysis of the central regions, where the sensitivity to background variations is
not as large as in the outskirts.

\subsection{Spectral analysis} \label{sec: sp ana}
To investigate cluster properties in the central regions, we accumulated spectra in two regions:
the inner region is  a circle of radius 0.05~$R_{180}$ and the outer region
an annulus with bounding radii 0.05-0.20~$R_{180}$ (the outer radius of the latter region
is limited by the apparent size of the nearest clusters).
Both IN and OUT regions are centered on the X-ray emission peak.
We recall that $R_{180}$ is the radius within which the mean density is
180 times the critical density and that it has been computed as in \citet{leccardi08b}.
Since, as discussed in Sect.~\ref{sec: entropy},  $R_{180}$ is itself computed from the temperature measured in
the OUT region we have iterated the process until it converged to stable values of  $R_{180}$,
the first guess for $R_{180}$ was computed using temperatures  from \citet{leccardi08b} and the
literature.

For each EPIC instrument and each region, we generated an effective
area (ARF), and for each observation we generated redistribution functions (RMF) for MOS1, MOS2,
and pn.

Spectra accumulated in the central regions of clusters have almost always high statistical
quality; therefore, the complicated procedures we developed for dealing with the background
in the outer regions (see Appendices of \citealt{leccardi08b}) are not strictly necessary and also EPIC
pn data have been used.
For all three detectors (namely MOS1, MOS2, and pn), channels were assembled in order to have
at least 25 counts for each group, as commonly done when using the $\chi^2$ statistic.
We merged nine blank-field observations to accumulate background spectra.
For each cluster observation, we calculated the count rate ratio, $Q$, between source and
background observations above 9~keV in an external ring (10$^\prime$--12$^\prime$) of the
field of view.
We scaled the background spectrum by $Q$ and, for each region, subtracted it from the
corresponding spectrum from cluster observation.
This rough rescaling accounts for possible temporal variations of the instrumental background
dominating at high energies without introducing substantial distortions to the source spectrum in the
soft energy band where cosmic background components are more important and the source 
outshines the background by more than one order of magnitude.

The spectral fitting was performed in the 0.5-10.0~keV energy band using the $\chi^2$
statistic, with an absorbed thermal model (WABS*MEKAL in XSPEC
v11.3\footnote{http://heasarc.nasa.gov/docs/xanadu/xspec/xspec11/index.html}).
We fit spectra leaving temperature and normalization free to vary.
The metallicity\footnote{The solar abundances were taken from \cite{anders89}.} was
constrained between $\pm5 \; Z_\odot$ (see the discussion in Appendix~A of \citealt{leccardi08a}).
The redshift was constrained between $\pm$5\% of the optical measurement, and the equivalent
hydrogen column density along the line of sight, $N_\mathrm{H}$, was fixed to the 21~cm
measurement \citep{dickey90}.
Finally, for each quantity we computed the average over the three (MOS1, MOS2, pn) values and
derived the projected emission measure, $E\!M$, as the ratio between the normalization and the
area of the region expressed in square arcminutes.

\section{Defining interesting quantities} \label{sec: entropy}
To characterize temperature and emission measure gradients, we need to compare the
central to a global value for such quantities; however, for local clusters it was only
possible to perform reliable measurements out to a small ($\approx$ 20\%) fraction of
$R_{180}$.
As a temperature reference, we used $kT_\mathrm{OUT}$ (see Sect.~\ref{sec: sp ana}), which
is found to be a good proxy for the global temperature from Fig.~4 of \citet{leccardi08b}.
\begin{figure}
  \centering
  \hspace{-1cm}\resizebox{9cm}{!}{\includegraphics{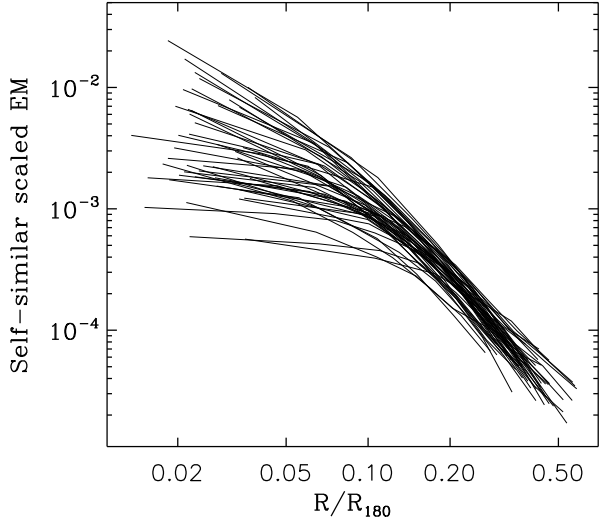}} \\
  \caption{Self-similar scaled emission measure profiles for the
  intermediate-redshift sample presented in \citet{leccardi08a,leccardi08b}. The radius is in units of
  $R_{180}$ and the emission measure is scaled by $E\!M_\mathrm{SSS}$ (see
  Eq.~\ref{eq: EM_SSS}).}
  \label{fig: EM_prof}
\end{figure}
A reference value for the emission measure should be measured at $\approx$~0.4~$R_{180}$,
where profiles show a remarkable degree of similarity (see Fig.~\ref{fig: EM_prof}).
 We have considered two different proxies for the emission measure at large 
radii finding that one of them, namely the emission measure calculated in the  
outer region, $E\!M_\mathrm{OUT}$, does somewhat better than the other. 
We have adopted $E\!M_\mathrm{OUT}$ as a proxy for the emission measure at large radii; 
the interested reader may find a detailed analysis of how well $E\!M_0$  and the other proxy 
approximate the emission measure at large radii in  App.~A.
For the less technically minded readers suffice it to say that $E\!M_0$ reproduces
the emission measure around 0.4~$R_{180}$ to better than 20\%.

\begin{figure}
  \centering
  \hspace{-0.82 truecm}\resizebox{9.8 truecm}{!}{\includegraphics{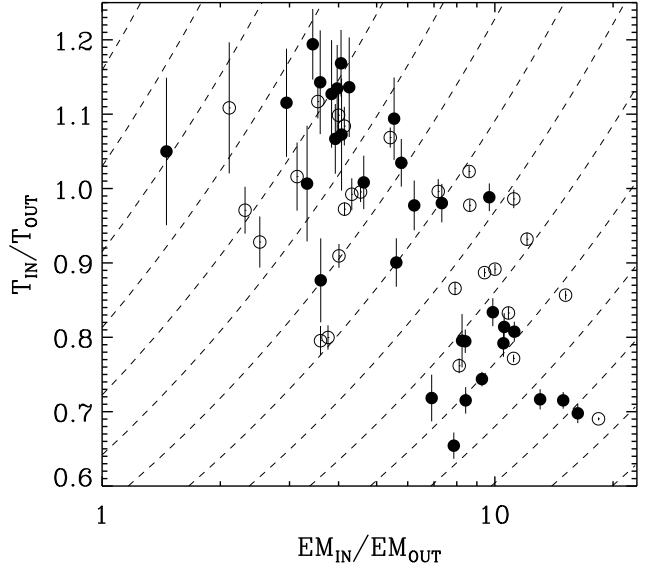}} \\
  \caption{Comparison of temperature and emission measure ratios for all clusters. The dashed
  curves indicate the regions of the plot where the pseudo-entropy ratio,
  $\sigma$, defined in Eq.(\ref{eq: sigma}), is constant; clusters with the strongest pseudo-entropy 
 variations populate the
  bottom-right corner. Open and filled circles indicate local (i.e. $z~<$~0.1) and distant
  (i.e. $z > 0.1$) clusters respectively.}
  \label{fig: SB_T plot}
\end{figure}
In Fig.~\ref{fig: SB_T plot} we compare the temperature ratio, $T_\mathrm{IN}/T_\mathrm{OUT}$,
with the emission measure ratio, $E\!M_\mathrm{IN}/E\!M_\mathrm{OUT}$.
As expected, there is a clear, but quite scattered, correlation: more precisely, the stronger
the emission measure peak, the stronger the temperature drop.
A key thermo-dynamic observable in describing clusters is  the entropy
\citep[e.g.][]{ponman03,voit05,pratt06}, which is commonly defined as:
$S \equiv T_X \times n_e^{-2/3}$, where $T_X$ and $n_e$ are the deprojected temperature and electron density.
In the literature \citep[e.g.][]{rossetti07}, it is common to define a
pseudo-entropy from projected quantities as: $s \equiv T \times E\!M^{-1/3}$.
Since we are interested in comparing core with cluster properties we
use our emission measure and temperature ratios to define a pseudo-entropy ratio,
\begin{equation}
\sigma \equiv (T_\mathrm{IN}/T_\mathrm{OUT}) \times (E\!M_\mathrm{IN}/E\!M_\mathrm{OUT})^{-1/3} .
\label{eq: sigma}
\end{equation}
The pseudo-entropy ratio has been found to be well correlated with the entropy ratio,
which is computed using deprojected quantities (see \citealt{Ross09} for a detailed comparison).
In Fig.~\ref{fig: SB_T plot} the dashed curves indicate the regions where
$\sigma$ is constant; clusters with the strongest variations (i.e. lower ratios) of
pseudo-entropy fill the bottom-right corner and are usually known as cool core clusters
(see~Sects.~\ref{sec: class sch} and \ref{sec: entr vs alt} for a detailed discussion).
Values of the pseudo-entropy ratio for all objects in our sample are reported in
Table~\ref{tab: sigma_z}.

The central cooling time is another quantity largely used in the literature
\citep[e.g.][]{peres98} to estimate the degree of relaxation of clusters.
As done for the entropy, we defined a pseudo-cooling-time,
$t_{pc} \equiv T^{1/2} \times E\!M^{-1/2}$, and a pseudo-cooling-time ratio,
\begin{equation}
\tau \equiv (T_\mathrm{IN}/T_\mathrm{OUT})^{1/2} \times (E\!M_\mathrm{IN}/E\!M_\mathrm{OUT})^{-1/2}.
\label{eq: tau}
\end{equation}
When ordering clusters according to $\tau$, we find essentially the same results as
when using $\sigma$.

An interesting property of our pseudo-entropy ratio is that it can be constructed from data of moderate statistical
quality  such as  serendipitous observations of clusters in deep \emph{XMM-Newton} and \emph{Chandra}
observations.
In Sect.~\ref{sec: compare core} we will employ $\sigma$ to divide clusters into 3 broad categories
namely: low entropy core (LEC), medium entropy core (MEC) and high entropy core (HEC) systems.
Before we proceed with our entropy based classification we must first consider other
classification schemes based, at least in part, on different cluster properties.

\section{Alternative classification schemes} \label{sec: class sch}

We wish to provide alternative classification schemes with which to compare our entropy based scheme.
More specifically we wish to compare with: 1) the traditional cool core/non-cool
core classification; 2) a classification scheme based on dynamical
properties.
To avoid circularity, we want the alternative 
classification schemes to be as independent as possible from our
measurements.
Therefore, for what concerns the cool core/non cool core classification, we 
rely on how astronomers have classified these clusters in the literature 
and not on our own data. More specifically,
we divide objects in 3 classes:
cool cores (CC), intermediate systems (INT) and non-cool cores (NCC).
We classify as CC those systems for which we find evidence in the literature of a temperature
decrement and a peaked surface brightness profile. NCC are those that possess neither of the above properties,
while INT systems are those that possess one or the other or alternatively both but not as well
developed as in full blown cool core systems.
 We refrain from providing a more quantitative classification for two main reasons: 1) this would be rather difficult to derive from the literature, indeed different authors make use of somewhat different criteria and 
certainly do not analyze data in a homogenous fashion; 2) our own entropy classification, as we shall 
see in Sect.~\ref{sec: compare core}, is best understood as a more quantitative classification scheme 
along these very lines. 

As far as the dynamical classification is concerned we divide object in 2 classes:
mergers (MRG), and systems for which we do not find evidence of a merger (NOM).
 We consider evidence for substantial cluster-wide interaction
leading to a merger classification the following phenomena:
1) cluster wide diffuse radio emission such as radio haloes and relics;
2) multi-peaked velocity distribution from optical spectroscopy;
   or evidence of substructure from the combination of optical spectroscopy
   and photometry; or evidence for multiple mass peaks from lensing analysis;
3) significant irregularities observed in X-rays both in  morphology and temperature maps.
 The lack of diffuse radio emission or of  multi-peaked velocity distribution
from optical spectroscopy is in itself insufficient to classify an object as relaxed.
Similarly  the absence of significant substructure on cluster wide scales in X-ray images
is not in itself proof that substructure does not exist or that the system under scrutiny is relaxed.
For these reason it is rather difficult to classify a system as relaxed, what we can
ascertain is that some systems lack evidence of merger activity. Consequently we  classify all objects
for which we do not have evidence for merging as ``no observed merging'' (NOM) systems.
We reiterate that a NOM system is not necessarily a relaxed system  but rather
a system for which we do not have evidence of  merging activity, in the sense
described previously in this paragraph.

In Table~\ref{tab: classif} we provide results from our classification work.
Columns 2 and 3 refer to radio emission, in the first column we indicate
with ``H'' clusters with radio haloes, with ``N'' clusters without radio haloes,
with ``?'' clusters with tentative radio haloes, with ``M'' clusters with mini-radio haloes and
with ``R'' clusters with radio relics. Column 3  provides  references  for column 2.
Columns 4 and 5 refer to optical emission,  we indicate
with ``Y'' clusters with substructure in the forms described above, with
``N'' clusters without and with ``?'' uncertain cases. Column 5 provides references for column 4.
Columns 6 and 7 refer to X-ray emission, in column 6 we indicate with ``CC''
clusters which have been identified as cool core (or cooling flow) systems;
with ``NCC'' clusters that have been identified as non-cool core (or non-cooling flow)
systems; with ``INT'' clusters with intermediate cores and
with ``MRG'' clusters identified as mergers. Column 7 provides references for column 6.
In Column 9 we provide our core based
classification and in column 10 our dynamical classification.
For all objects where a classification cannot be desumed directly from information in
columns 2 to 7 we provide a note explaining how the classification was derived.
In column 8 we indicate those objects for which we provide a note with ``Y'' and
those for which we do not with ``N''.

As far as the core classification is concerned, we classify 24 systems as CC, 25 as NCC and 10 as
INT.
As far as the dynamical classification is concerned we classify  19 objects as MRG and 40 as
NOM.

\subsection{Notes on individual objects}

\subsubsection{A4038}
{\it Core Classification }

\noindent
Analysis of \emph{Chandra} data provides a central cooling time of 1.3 Gyr \citep{Su07} and a flat
temperature profile \citep{Sn06}, moreover inspection of \emph{Chandra}
and \emph{XMM-Newton} images show  evidence of an irregular core. This system clearly does not host a full blown cool
core, we classify it as intermediate.

\noindent
{\it Dynamical Classification}

\noindent
Diffuse radio emission has been reported for this source \citep{Sl98,Sl01},
although the emission is located at the center of the cluster, its appearance is more
similar to that of a radio relic than to a radio halo: it is most likely a remnant
associated to the radio galaxy observed in this system.
Optical observations  \citep{Bu04} provide controversial evidence for substructure on large scales.
The evidence pointing to a merger are in our opinion insufficient, we  choose to classify this object as NOM.

\subsubsection{A3571}
{\it Core Classification }

\noindent
Analysis of \emph{Chandra} data provides a central cooling time of 1.3 Gyr \citep{Su07}, however there is
no evidence for a temperature decrement in the core \citep{Sa06}.
This system does not host a full blown cool core, we classify it as INT.

\noindent
{\it Dynamical Classification}

\noindent
On the basis of the multi-wavelength properties of the A3571 cluster complex,  \citet{Ve02} propose that
A3571 is a very advanced merger, and explain the radio properties derived from their study in the light
of this hypothesis. We deem the evidence collected by \citet{Ve02} insufficient to classify A3571
as a merger and conservatively catalog it as NOM.

\subsubsection{A1650}
{\it Core Classification }

\noindent
Analysis of \emph{Chandra} data, readily available through the ACCEPT archive \citep{Ca09}, show this object
to be of an intermediate nature possessing many of the traits typical of cool cores such as an abundance excess
and a  temperature drop, albeit with a relatively high core entropy $\simeq$ 40 keV cm$^2$.  \citet{Do05} define A1650 as
a radio quiet cool core speculating that the entropy has been augmented by a recent AGN triggered heating
event also responsible for halting the AGN feeding process and ensuing radio manifestations. We classify A1650 as intermediate.

\subsubsection{A1689}
{\it Core Classification }

\noindent
Analysis of \emph{Chandra} observations by \citet{Ca09}
show evidence for a well defined core with a metal abundance excess and a relatively high
central entropy of about 80 keV cm$^2$.
We classify A1689 as an intermediate system.

\noindent
{\it Dynamical Classification}

\noindent
Optical studies find evidence for two velocity peaks, possibly due to line of sight superposition \citep{Gi01}.
This was later confirmed by \citet{Lo06} who performed a detail  kinematic study  of about 200 galaxies with
measured redshifts; \citet{An04} find circumstantial evidence for a merger.
We deem the evidence  insufficient to classify A1689
as a merger and conservatively catalog it as NOM.

\subsubsection{A963}
{\it Core Classification }

\noindent
Analysis of \emph{Chandra} observations by \citet{Ca09}
show evidence for a well defined core with a  modest temperature decrement, a metal abundance excess
and a relatively high central entropy of about 60 keV cm$^2$.
We classify A963 as an intermediate system.


\section{Entropy vs. alternative classification schemes} \label{sec: entr vs alt}

In this section we compare our entropy classification scheme with the core and dynamical
classification schemes presented in the previous section.

\subsection{Entropy vs. cool core classification scheme } \label{sec: compare core}
\begin{figure}[!ht]
  \centering
  \hspace{-0.82 truecm}\resizebox{9.8 truecm}{!}{\includegraphics{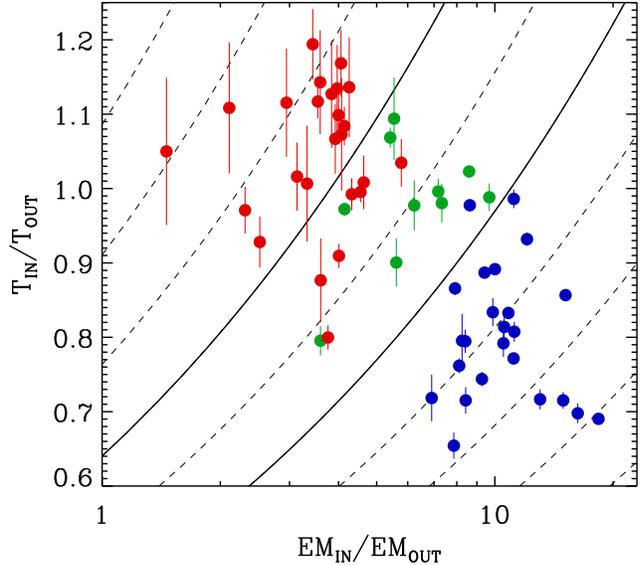}} \\
  \caption{Comparison of temperature and emission measure ratios for all clusters. The dashed
  and the solid curves indicate the regions of the plot where the pseudo-entropy ratio,
  $\sigma$, is constant; clusters with the strongest pseudo-entropy variations populate the
  bottom-right corner. The solid curves indicate the thresholds used to divide clusters in
  high (HEC), medium (MEC) and low (LEC) entropy core systems.
  Colors are associated to the cool core classification scheme, namely non-cool cores
  (NCC) are red, intermediate (INT) are green and cool cores (CC) are blue.
}
  \label{fig: SB_T_alter plot}
\end{figure}

In Fig.~\ref{fig: SB_T_alter plot}  we compare our entropy (Sect.~\ref{sec: entropy}) and cool core
(Sect.~\ref{sec: class sch}) classification schemes, we do this by
plotting the temperature versus the emission measure ratio as in Fig.~\ref{fig: SB_T plot}
with the nuance that we use colors
to differentiate objects belonging to different classes, namely we use
red for non-cool core (NCC) systems, green for intermediate (INT) systems and blue for cool cores (CC).
 \setcounter{table}{3}
 \begin{table}[!ht]
\
  \caption{Summary of classification schemes.}
  \label{tab: class}
\begin{tabular}{|l|l|l|}
\hline
Scheme & Classification & Acronym \\ \hline
\multirow{3}{*}{Entropy Ratio}      & Low Entropy Core    & LEC \\
                                    & Medium Entropy Core & MEC \\
                                    & High Entropy Core   & HEC \\
\hline
\multirow{3}{*}{Core Properties}    & Cool Core           & CC  \\
                                    & Intermediate        & INT \\
                                    & Non Cool Core       & NCC \\
\hline
\multirow{2}{*}{Dynamic Properties} & No Observed Merging & NOM \\
                                    & Merging System      & MRG \\
\hline
\end{tabular}
\end{table}
Inspection of Fig.~\ref{fig: SB_T_alter plot} shows that, as expected, there is some 
correlation between the two classifications. More specifically we find that:
1) cool cores have small values of $\sigma$ or, in other words, are
characterized by a strong pseudo-entropy gradient;
2) intermediate objects have intermediate entropy gradients;
3) non-cool core, on average, have small entropy gradients.
We employ  Fig.~\ref{fig: SB_T_alter plot} to divide our objects in
3 broad entropy classes namely:
low entropy core (LEC) systems, medium entropy core (MEC) systems
and high entropy core (HEC) systems.
Given the continuous distribution of objects the precise values of  $\sigma$ adopted to separate
LEC from MEC and MEC from HEC are of course somewhat arbitrary.
One possible criterion is that all INT objects belong to the MEC class.
By adopting such a criterion we set the separation between LEC and MEC at $\sigma = 0.45 $
and the separation between MEC and HEC at $\sigma = 0.64 $. In Table~\ref{tab: sigma_z}
we report the pseudo-entropy ratio and the entropy class for all objects in our sample.
To help our readers navigate through the three different classification schemes we have presented,
we provide in Table~\ref{tab: class}  a brief summary including the acronyms
that are used extensively in this paper.

Interestingly, while the CC and INT systems separate out quite
well in terms of their entropy ratios, the intermediate and
non-cool core systems appear to be more mixed up.
We find that only 1 CC systems is
classified as a MEC and that 7 NCC systems are classified as MEC.
The excellent match between  LEC systems and cool cores is by no means a surprise,
indeed one of the possible definitions of a cool core cluster is that of a system hosting
a low entropy core \citep[e.g.][]{Ca09}. The agreement should rather be viewed as yet another demonstration of the
effectiveness of the $\sigma$ indicator in describing the entropy profiles
of clusters. To a lesser extent the same argument may be applied to the MEC vs. INT systems comparison
and to the HEC and NCC comparison, however for these systems, particularly for the latter,
it becomes progressively more difficult to define the core and its properties. Indeed the more
attentive amongst our readers may recall that in Sect.~\ref{sec: sample} we refrained from including a
number of clusters with poorly defined cores in our sample for the very reason that they would be difficult
to classify.

\begin{figure}
  \centering
  \hspace{-0.82 truecm}\resizebox{9.8 truecm}{!}{\includegraphics{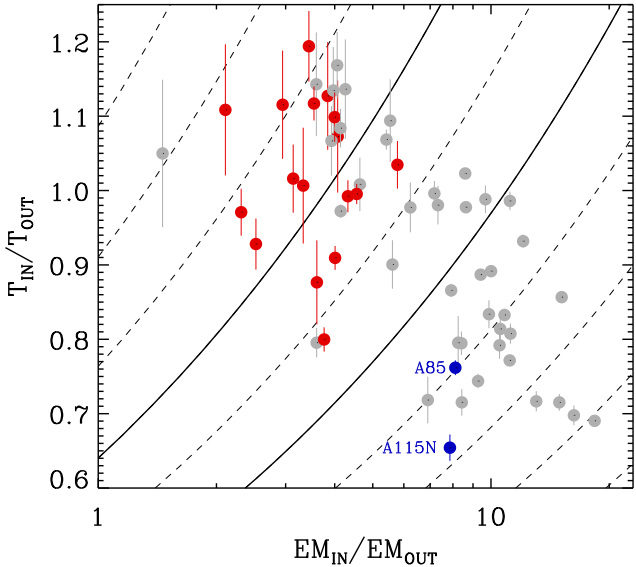}} \\
  \caption{Comparison of temperature and emission measure ratios for all clusters. The dashed
  and the solid curves indicate the regions of the plot where the pseudo-entropy ratio,
  $\sigma$, is constant; clusters with the strongest pseudo-entropy variations populate the
  bottom-right corner. The solid curves indicate the thresholds used to divide clusters in
  high (HEC), medium (MEC) and low (LEC) entropy core systems.
  Colors are associated to the dynamics based classification scheme, namely mergers
  (MRG) are red, with the exception of A115N and A85 which are blue,
  and  systems that do not show evidence for merging (NOM) are gray.
}
  \label{fig: SB_T_mrg}
\end{figure}

\subsection{Entropy vs. dynamical classification scheme } \label{sec: compare dynamical}

In Fig.~\ref{fig: SB_T_mrg}  we compare our entropy (Sect.~\ref{sec: entropy}) and dynamical
(Sect.~\ref{sec: class sch}) classification schemes, we do this as in Fig.~\ref{fig: SB_T_alter plot}
with the difference that the color coding now refers to the dynamical classification,
more specifically
red for merging (MRG) systems, with the exception of A115N and A85 which are blue, and gray 
for systems that do not show evidence for merging (NOM).
Inspection of Fig.~\ref{fig: SB_T_mrg} shows that there is some correlation between
the two classifications. More specifically we find that:
1) the  majority of mergers are HEC systems, a few  are classified as MEC systems and 2 are LEC systems;
2) NOM systems are found all over the plot.
The result on NOM systems has a trivial explanation,  as already discussed in Sect.~\ref{sec: class sch},
these are objects for which we do  not observe evidence of merging:
they will include both systems that are not undergoing a merger and
mergers for which we do not have observational evidence of merging activity.
A more interesting result is the one on objects identified as mergers.
As already noted only a few mergers are MEC and LEC, let us now focus on the 2 MRG
with the lowest pseudo-entropy ratios, these are A85 and A115N, which are both LEC systems.
A85 and A115N, which are plotted in blue in  Fig.~\ref{fig: SB_T_mrg}, are systems where X-ray observations clearly show the presence of  two clumps.
In both cases the evidence found in the literature
supports a scenario where the effects of the
merger have not reached the core of the main structure (which is the one for which we computed the entropy
ratio), either because the merger is an off-axis merger, A115N \citep{Gu05}, or
because it is in an early stage, A85 \citep{Ke02} and  A115N \citep{Ba07b}.
A potential concern for these objects is that, if the sole evidence for the merger not
having reached the core were the presence of the core itself, the whole argument would of course be circular
and not particularly convincing.
We note that if the mergers were in an advanced state we would expect
distorted morphology and irregular  temperature distribution in the
circum-core regions, as well as substantial displacement between X-ray and optical light peaks;
this is indeed what is observed in other merging systems (e.g A2256, \citealt{sun02} and \citealt{bourdin08}; A3667,
\citealt{briel04} and \citealt{vikh01}) and  predicted in
simulations  \citep[e.g.][]{rowley04,ric01}.
In the cases of A85 no such evidence is found.  In the case of A115N, \citet{Gu05}
find evidence for heating of the region separating the cores, but no indication of supersonic motion,
moreover, \citet{Ba07b} detect
 two optical substructures  of cluster-type well recognizable in the plane of the sky
and roughly coincident with the X-ray peaks thereby favoring a pre-merger scenario.
Therefore, in  A115N, we have evidence of some form of interaction of minor intensity, that may
be explained either in the context of an off-axis merger or in that of an early stage of
the merger process.

It is quite interesting that when comparing our core entropy based classification with
the dynamical classification, the only 2 mergers to possess a LEC are systems
where the effects of the merger have not reached the core. The rather obvious inference,
which will be discussed at some length further on in the paper (see Sect.~\ref{sec: Z vs S}), is that
mergers do have the capability of disrupting low entropy cores.
For the time being we note that, if we exclude those interacting systems for which the effects of
the merger have not reached the core, we find that MRG systems have pseudo-entropy ratios larger than 0.51.
A potential concern is that the same observational evidence may have been used to classify an object
as a  merger and as a HEC, this would of course provide a rather trivial explanation for the
correlation between the two classifications. We note that the presence of a well defined core does not
imply that an object may not also show substructure in its surface brightness and temperature
maps, indeed A1644 \citep{Re04} and A115N \citep{Gu05} are both good examples of such systems.
Moreover only 3 out of the 17 bone-fide mergers, (we have excluded the 2 special cases of A85 and A115N)
have been classified as mergers on the basis of their X-ray properties alone; for the
other 14 systems there is evidence for a merger from radio and/or optical observations.

In summary the comparison of our entropy and dynamics based classification schemes shows that dynamically active
systems tend to have high entropy cores while, with the exception of A85 and A115N where the effects of
the mergers have not reached the core, low entropy cores are not found in merging systems.

\subsection{Comparison with previous work} \label{sec: compare}

Ours is not the first attempt to divide clusters on the basis of  their core properties.
There have been various works concentrating on somewhat different core properties:
\citet{Sn06}, for instance, consider the core temperature as  discriminator, they  define
as cool core clusters  those systems for which the ratio between average cluster and core temperature
exceeds unity at greater than 3$\sigma$ significance. The average cluster temperature is determined from
an annulus with bounding radii 0.1-0.2 $R_{500}$ and the core temperature from a circle with radius
0.1~$R_{500}$.
The circle is similar to our inner region while the annulus is somewhat smaller than our outer region.
While the selection procedure appears to work  well for the specific objects
in the \citet{Sn06} sample, it has some rather obvious pitfalls,
visual inspection of Fig.~\ref{fig: SB_T_alter plot} shows that the range
$ 0.8 < T_\mathrm{IN}/T_\mathrm{OUT} < 1.0 $
is populated by objects belonging to all 3 entropy classes, i.e. LEC, MEC and HEC,  it is only the
additional use of the
$E\!M_\mathrm{IN}/E\!M_\mathrm{OUT}$ ratio that allows us to provide a more effective means of separation.
As an example of the limitations associated to a classification system based on the temperature 
decrement alone, we may consider Fig.~\ref{fig: SB_T_mrg} where we observe that, contrary to what is 
found when employing the entropy classification scheme, a sizeable fraction of mergers are found in 
clusters  without temperture decrement.

\begin{figure}
  \centering
  \hspace{-0.82 truecm}\resizebox{9.8 truecm}{!}{\includegraphics{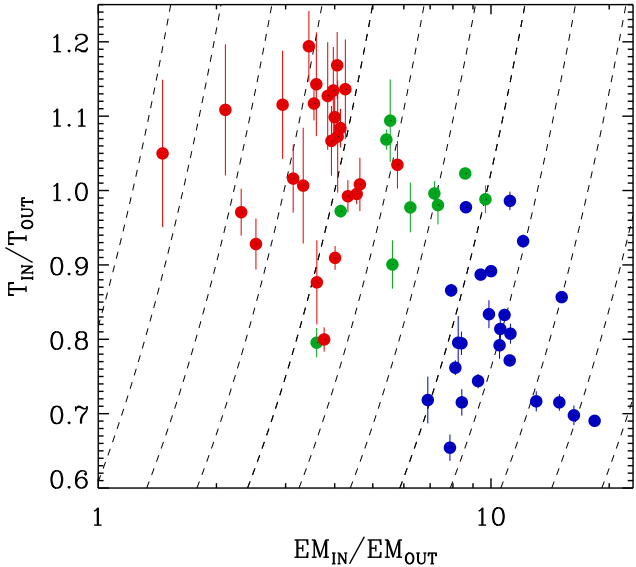}} \\
  \caption{Comparison of temperature and emission measure ratios for all clusters. The dashed
  curves indicate the regions of the plot where the pseudo-cooling-time ratio,
  $\tau$, defined in Eq.~\ref{eq: tau}, is constant; clusters with the strongest pseudo-
  cooling-time variations populate the bottom-right corner.
  Colors are associated to the cool core classification scheme, namely non-cool core
  (NCC) are red, intermediate (INT) are green and cool cores (CC) are blue.
}
  \label{fig: SB_T_tcool plot}
\end{figure}

\citet{baldi07} use the cooling time, or better the ratio of cooling time to age of the universe
at the cluster redshift. As already noted in Sect.~\ref{sec: entropy}, a pseudo-cooling-time ratio, $\tau$
defined as in Eq.~\ref{eq: tau}, separates out clusters in much the same way the pseudo-entropy-ratio does.
This is illustrated in Fig.~\ref{fig: SB_T_tcool plot}  where we show the same plot reported in
Fig.~\ref{fig: SB_T_alter plot} with the only difference that the dashed lines indicate region
of constant $\tau$ rather than $\sigma$.

Other authors have compared a core based classification with a dynamics based
classification.
\citet{Mc04} divide objects in cooling flow and non-cooling flow depending on the presence
or absence of a temperature gradient in their cores, this is essentially the same classification
adopted by \citet{Sn06}, they also provide a dynamical classification dividing their objects
in relaxed and non-relaxed on the basis of the presence or absence of large scale (a few hundred
kpc) substructure in the X-ray images, presumably related to mergers. They find that, of the 33 objects
in their sample,  18 are cooling-flow and 15 non-cooling flow. Interestingly, of the 18 cooling flow objects,
12 are classified as relaxed and 6 as non-relaxed and of the 15 non-cooling flow objects 9 are non-relaxed
and 6 are relaxed. The presence of a sizable number of non-relaxed cooling flow systems and of relaxed
non-cooling flow systems is considered as evidence for a different origin of cooling flows and the relaxed/non-relaxed
state in clusters. \citet{Mc04} propose a scenario where an object ends up being cooling or non-cooling flow
on the basis of the amount of entropy injected in the system, while the relaxed or non-relaxed
nature depends on the object having or not having recently experienced a merger.
Our own findings appear to be at variance with what has been reported by \citet{Mc04}: the objects we classify as
mergers, with the exception of systems where the effects of the merger have not reached the core, are all
characterized by a high pseudo-entropy ratio.
Since 15 of the 33 object in McCarthy's sample are also found in ours we have compared our results with
theirs \footnote{In this comparison we make use of our dynamical and entropy classifications only,
the cool core classification is omitted as it provides results which are identical to those
derived from the entropy classification.}.
For 11 of the 15 objects that are in common, their classification
is in agreement with ours, more specifically:
8 objects are classified as cooling flow and relaxed by  \citet{Mc04} and as LEC and NOM by ourselves;
2 objects are classified as non-cooling flow and non relaxed by  \citet{Mc04} and as non-low entropy systems,
either HEC and MEC, and mergers by ourselves; one object, namely A115N, is classified as
cooling flow and non-relaxed by  \citet{Mc04}  and as LEC and MRG by ourselves.
For 4 objects their classification appears to differ from ours.
More specifically there is  one object,
namely A1068, which we classify as LEC and non-merging and \citet{Mc04} classify as
cooling flow and non-relaxed. \citet{Mc04} refer to a paper by \citet{Wi04}, which indeed provides evidence for
substructure in surface brightness, temperature and metal abundance, however all images presented in that paper
cover a region of 200~kpc~x~200~kpc and therefore include only the cool core. The kind of substructure found
in the core of A1068 is akin to that found in many other cool core systems and is generally believed to be
associated to the AGN found at the center of these systems and not to a merger.
A second object, A85, is classified as LEC and MRG by ourselves and as non-cooling flow and
non-relaxed by \citet{Mc04}, however the authors do not specify if the core property is referred to
the main structure, which is a well known cool core \citep[e.g.][]{peres98} or to the sub-structure which hosts an
intermediate core \citep{Ke02}, assuming the latter is the case than the difference in core classification
is trivial as we refer to the main structure.
The last 2 objects, A1413 and A1689, are  classified by \citet{Mc04} as non-cooling flow and relaxed
while we classify them as MEC and NOM.
A first important point is that \citet{Mc04}  define as relaxed those objects for which they do not find evidence
of large scale irregularities in the X-ray images, which however does not necessarily imply that these objects are
indeed relaxed (an example of such relatively rare systems is A401, a cluster with fairly regular X-ray morphology, \citealt{Sa04},
featuring a small radio halo, \citealt{Ba03}, and significant structure in X-ray temperature, \citealt{Sa04} and
\citealt{bourdin08}).
We classify A1413 and A1689
as  MEC  in our entropy classification system;
both these systems possess a well defined core which, however, is not as
prominent as the ones typically found in LEC systems.
Indeed analysis of \emph{Chandra} observations of A1413 and A1689
provide evidence for a well defined core with a modest temperature drop in the core \citep{vikh05,Ca09},
a metal abundance excess \citep{vikh05,Ca09} and a relatively high central entropy of about 60 keV cm$^2$
and 80 keV cm$^2$ respectively \citep{Ca09}.

In summary by comparing our classification schemes with those provided by  \citet{Mc04} we find
that, of the 4 objects for which we do not agree, the 3 objects with mixed classifications i.e. mergers with cooling flows (A1068) and relaxed non-cooling flows
 (A1413 and A1689) cannot be used to  support a
scenario where the absence or presence of a cooling flow is
unrelated to the object having or not having experienced a merger.
Indeed the former object is non-relaxed on small scales in much the same way as other cool core systems and
for the latter two: I) the lack of evidence for a merger does not necessarily mean that a merger is not present;
II) the classification as non-cooling flow is insufficiently accurate, as both these system host well defined
intermediate cores.

\section{Chemical properties} \label{sec: Z scheme}
In this section we compare chemical  with thermo-dynamic properties for the objects in our sample. As
a first step we discuss metal abundance profiles.

\subsection{Metallicity profiles} \label{sec: Z prof}

We have divided our sample in the entropy classes defined in Sect.~\ref{sec: entr vs alt}
and produced mean radial metallicity profiles for each entropy class. Metal abundance profiles for
individual systems come from \citet{leccardi08a} for the intermediate redshift subsample and from \citet{Ross09} from the low redshift sample.
\begin{figure}
  \centering
  \hspace{-0.82 truecm}\resizebox{9.8 truecm}{!}{\includegraphics[angle=0]{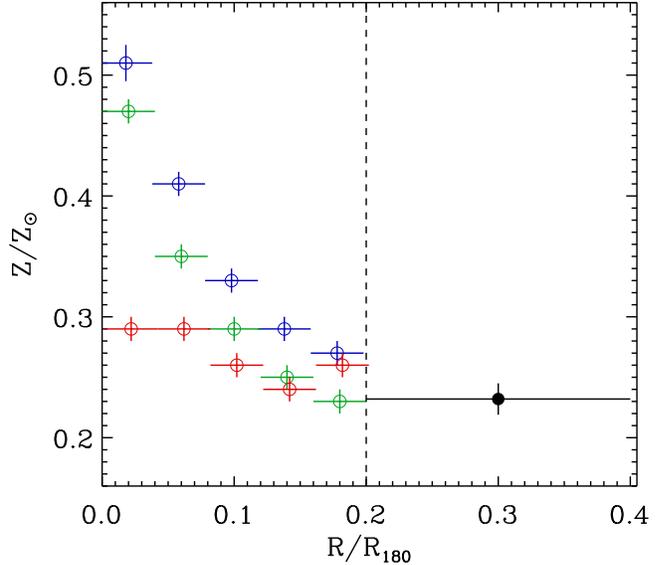}} \\
  \caption{Mean metallicity profiles for LEC (blue circles), MEC (green), and HEC (red)
  clusters. Abundances are expressed in \cite{anders89} solar values.
  The dashed line indicates 0.2~$R_{180}$, the radius within which profiles are
  obtained from all clusters, the black data point beyond 0.2~$R_{180}$ is the average value
  from distant (i.e. $z > 0.1$) clusters belonging to all entropy classes.}
  \label{fig: Z prof}
\end{figure}
In Fig.~\ref{fig: Z prof} we show the mean profiles for LEC, MEC, and HEC clusters,
the binning is in units of $R_{180}$ and was computed as in \citet{leccardi08a}; for each
bin the average is calculated by allowing for an intrinsic dispersion.
As can be seen in Fig.~\ref{fig: Z prof},
within 0.1~$R_{180}$, a region typically associated with the core, all profiles show
an abundance excess.
The excess is strongest for LEC, somewhat weaker for MEC, and weakest for HEC clusters.
Interestingly the modest excess observed in HEC clusters is  similar to the
one obtained from a \emph{BeppoSAX} sample of non-cool core clusters \citep{grandi01}
dominated by well known merging systems.
Between 0.1 and 0.2~$R_{180}$ the profiles for the three classes are roughly consistent with
one another and, at least for LEC and MEC systems, show a significant excess with respect to
the mean value measured in the outskirts.
Between 0.2 and 0.4~$R_{180}$, where we have data for the intermediate-redshift sample only,
the three profiles are consistent with being flat and equal with each other.
We therefore plot the abundance averaged over all 3 entropy classes, which turns out to be
$0.23 \pm 0.01 \; Z_\odot$, and in good agreement with the value obtained by
\cite{grandi01} for a local sample of relaxed clusters observed with \emph{BeppoSAX}.
As already discussed in  Sect.~\ref{sec: entropy}, the precise choice of pseudo entropy ratio values
used to divide objects in the three entropy classes is rather arbitrary; by experimenting with slightly different
values we find that, while the details of the profiles may change somewhat as borderline objects
are shifted from one class to another, the qualitative description provided above remains valid.

An important point is that the abundance measures in the outer region are not only consistent
with being flat but also appear to be independent of
the entropy class (HEC, MEC or LEC) or of the dynamical class (MRG or NOM) of the object.
Moreover the mass of ICM enclosed within 0.2 and 0.4~$R_{180}$ is about two times that contained within
0.2~$R_{180}$ and, according to \cite{grandi04}, the Fe  mass in the abundance excess of 
CC clusters is roughly 10\% of the Fe mass integrated out to  0.25~$R_{180}$.
It follows that
estimates of how the global metal abundance varies with respect to other quantities
are best performed by making measures in the 0.2-0.4~$R_{180}$ range. An example, which we shall
 not discuss further, is the often quoted
anti-correlation between metal abundance and temperature \citep{baumgartner05,balestra07}.
Another example is the measure of the
evolution of the global metal abundance with cosmic time.
Current estimates \citep{balestra07,maughan08} are performed at small radii
where the presence of an abundance excess, more pronounced in some systems than in others,
poses a major obstacle.
Both \citet{balestra07} and \citet{maughan08} are  aware of these difficulties and
confront them either by gauging  how  the mix of cool
cores and non cool cores might affect the observed evolution in the iron
abundance \citep{balestra07}, or by excising the innermost region (0.15~$R_{500}$) from their spectra
 \citep{maughan08}.
A more robust approach would be to restrict measures to the 0.2-0.4~$R_{180}$ range.
In \citet{leccardi08a} we showed that by adopting the above radial range in the limited redshift interval covered by our
data, $0.1 < z < 0.3$, we could not discriminate between no variation of the abundance with redshift and a variation of the
kind described in \citet{balestra07}. Extension of these kind of measures out to $z\simeq$0.5, while
observationally challenging, would allow to discriminate between the two competing alternatives.

\subsection{Chemical vs. thermo-dynamic quantities} \label{sec: Z vs S}

\begin{figure}
  \centering
  \hspace{-0.82 truecm}\resizebox{9.8 truecm}{!}{\includegraphics{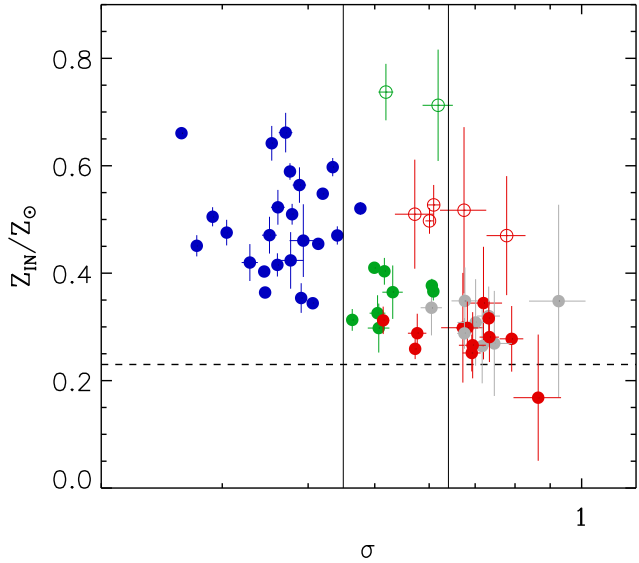}} \\
  \caption{Central metallicity, $Z_\mathrm{IN}$, vs.  pseudo-entropy ratio, $\sigma$.
   In blue and green we indicate objects that, within the core classification scheme, are
   respectively identified as  CC and INT systems. In red and grey we indicate objects that, within the
  dynamical classification scheme, are respectively identified as MRG and NOM. In the few instances
  where the two classification schemes require different colors, the core scheme takes precedence.
 Open circles indicate high metallicity MEC and HEC systems.
  Abundances are expressed in \cite{anders89} solar values. The dashed horizontal line at $0.23 \; Z_\odot$ indicates
  the mean metallicity in the outer regions of clusters \citep{leccardi08a}.
The solid vertical lines mark the boundary between the three entropy classes
(see Sect.~\ref{sec: compare core} for details).}
  \label{fig: Z_S plot}
\end{figure}
In Fig.~\ref{fig: Z_S plot} we plot the metallicity measured in our inner region, $Z_\mathrm{IN}$, vs. the
pseudo-entropy ratio, $\sigma$, for all clusters in our sample. Values of  $Z_\mathrm{IN}$
and $\sigma$ are also reported in Table~\ref{tab: sigma_z}.
We find $\sigma$ and $Z_\mathrm{IN}$ to have a negative correlation, namely: the stronger the
pseudo-entropy gradient, the stronger the metallicity peak.
This results tells that, baring  a few exceptions that we will discuss later in this section,
the metal rich gas in clusters also happens to be the low entropy gas.
An important point that we only mention here and that we have discussed in detail elsewhere
 is the presence of a consistent scatter in the abundance distribution, particularly for LEC
systems \citep{DeG09}.
The most interesting result is arguably the presence of a few (7) MEC and HEC systems
with unusually high metal abundance.
Of these high Z systems  4, namely A1644, AS0084, A576 and A3562, have metalicities well above the
typical values found for other MEC and HEC systems, while 3, namely A2034, A209 and  A1763, suffering from large
indetermination in there abundance estimates, show poor statistical evidence of an excess (roughly 2$\sigma$).
For A2034, a long \emph{Chandra} observation \citep{baldi07} provides a somewhat tighter constraint  which raises
the significance to about 3$\sigma$.

From our classification work (see Sect.~\ref{sec: class sch}) we know that a large fraction of these objects,
 5/7, are mergers (they are shown in red in Fig.~\ref{fig: Z_S plot}). In other words, most
of our high metallicity MEC and HEC systems  are undergoing a phase of rapid dynamical change.
This simple consideration leads to the
question of the original equilibrium configuration from which these systems evolved.
There are two issues that should be kept in mind when addressing this question.
The first is that metals are reliable markers of the ICM, in the sense that, once metals
have polluted a given region of a cluster, the timescale over which the same metals will diffuse
is comparable, likely longer, than
the Hubble time \citep{sar88,chuzhoy03}.
Thus, for all practical purposes, metals  trace the region of the ICM where they have been injected
and can be diluted only if the ICM itself undergoes mixing processes.
The second is that abundances such as those observed in our metal rich MEC and HEC systems are found in the
cores of LEC systems, indeed our own analysis (see Fig.~\ref{fig: Z_S plot}) shows that every LEC system has an excess
with respect to the metal abundance found in cluster outer regions.

Keeping the above considerations in mind, the most likely explanation is that our high abundance
MEC and HEC clusters originate from LEC systems that have undergone substantial heating.
While for one of our objects, namely A1644, the relatively modest entropy ratio
is not inconsistent with heating from the central AGN, as is observed in other intermediate systems such
as A1650 \citep{Do05}, for all other systems the required heating is beyond what can be provided
by the central AGN and must come from some other mechanism. Since all but one of the other 6 high metallicity objects show evidence of a merger, it seems reasonable to assume that the heating may indeed be provided
by the merger event.

This interpretation however clashes with claims from at least 2 groups conducting cluster simulations.
According to \citet{poole06,poole08}  and \citet{burns08}, once cool cores form it is extremely difficult to
disrupt them. The above authors suggest that the fate of a cluster (LEC or otherwise) is decided early on in
its history; if it is subject to an event that raises its entropy than it will likely not develop a cool core
and subsequent mergers will be effective in maintaining the high entropy state.
Conversely, if a cool core is formed early on, it will be very difficult to disrupt, i.e. subsequent
mergers will not destroy the entropy stratification.
In our opinion the scenario described in  \citet{poole06,poole08} and  \citet{burns08} suffers from
two major shortcomings, one on the observational side, the other on the theoretical side.
Let us consider the former;
if the presence/absence of cool cores is not related to the dynamical state of a cluster one would
expect to observe LEC in some merging clusters, and  HEC  in some relaxed systems.
In Sect.~\ref{sec: class sch} we have shown that:
1) having excluded A85 and A115N, where the effects of the merger have not reached the core,
for none of the remaining 21 LEC clusters do we find evidence that they are located in a merging system;
2) all our merging systems, baring the afore quoted exceptions of A85 and A115N, are MEC or HEC systems, with the vast
majority being HEC (11/17).
These findings are at variance with those reported by \citet{Mc04} who do identify a few non-relaxed
cooling-flow systems  and relaxed non cooling-flow systems.
In Sect.~\ref{sec: compare} we have compared our entropy and dynamical classification schemes
with those presented in \citet{Mc04} for the 15 objects that are present in both samples
finding that, of the 4 cases where our classifications do not agree, 3 likely result from misclassifications
by \citet{Mc04}, while the fourth probably has a trivial explanation.

The second shortcoming is related to the fact that current simulations cannot reproduce observed
cool cores, rather they produce something more akin to traditional cooling flows.
\citet{poole06,poole08}  correct for this by imposing initial conditions in the
core so as to reproduce observed cool cores, however merger events run on timescales longer than the
cooling time in cores. Indeed, as pointed out in \citet{poole06}, within 0.5 Gyr,  well before the clusters
begin interacting significantly, the central entropy profile reverts to the self similar power law shape.
Alternatively \citet{burns08} strive to reproduce observed cool cores
by introducing ad-hoc sub-grid recipes.
Thus, if simulations cannot provide a self consistent picture of cool cores, why should we
be compelled to trust simulations that tell us that cool cores survive mergers?

Recently \citet{sanderson09} have provided observational evidence favoring scenarios where
cluster mergers are capable of erasing cool cores. These authors have shown that in a sample of
65 objects, the X-ray/BCG projected offset correlates with
the gas density profile. Under the assumption that the offset serves to measure the dynamical
state of the cluster, their result implies that the cool core strength progressively
diminishes in more dynamically disrupted clusters. Such a trend
is expected if cluster mergers are capable of erasing cool
cores.


\section{Summary} \label{sec: concl}

The main results presented in this paper may be summarized as follows.
\begin{itemize}

\item{} We have constructed an indicator of the entropy of the core relative to
that of the cluster, the pseudo entropy ratio $\sigma$. Our indicator is robust,
in the sense that somewhat different choices of the quantities from which the ratio
is computed result in very similar values of $\sigma$. The indicator is also relatively
parsimonious, in the sense that it may be constructed from data of moderate statistical
quality.

\item{} The classification of clusters based on the entropy indicator improves
upon the traditional classification scheme based on the presence or absence of
a temperature drop in the core. Conversely classification schemes based on the
central cooling time appear to be essentially equivalent to ours.

\item{} A comparison between the entropy based classification scheme and
a classification scheme based on dynamical properties shows that the large majority
of merging systems are characterized by large entropy ratios. Only 2 of
our merging systems feature a low entropy core (LEC) and in both cases we were
able to establish, with reasonable certainty, that the effects of the merger have
not reached the core.
Our findings are at variance with those presented by \citet{Mc04} who do find evidence
of non relaxed cooling flow systems and relaxed non-cooling flow systems.
We have compared our entropy and dynamical classification schemes
with those in \citet{Mc04} for the 15 objects that are common to both samples
finding that the 3 cases where our classifications do not agree likely result from misclassifications
by \citet{Mc04}.

\item{} We find that mean abundance profiles for our 3 entropy classes, namely low entropy
core (LEC), medium entropy core MEC, and high entropy core (HEC), may be divided in 3 regions. In the
outer region, between 0.2 and 0.4~$R_{180}$, all 3 profiles are consistent with being flat and with one another.
In the core region, within 0.1~$R_{180}$, all classes feature an excess with respect to the mean value found in the outer regions.
The excess is strongest for LEC, somewhat weaker for MEC, and weakest for HEC clusters.
Between 0.1 and 0.2~$R_{180}$ the profiles for the three classes are roughly consistent with
one another and, at least for LEC and MEC systems, show a significant excess with respect to
the mean value measured in the outskirts.

\item{}
We find that objects with stronger pseudo-entropy gradients have  more pronounced metallicity peaks.
This results tells that, baring  a few exceptions, the gas that is more enriched in metals  also happens to be the one
featuring the lowest entropy.

\item{} We have identified a small number of medium and high  entropy core systems with a large central
metallicity.  The majority of these objects have been classified as mergers, i.e. as systems far from
their equilibrium configuration.  We surmise that these systems evolved from low entropy core clusters
that have experienced a heating event. We have examined  simulation based claims that conflict with our conjecture
finding they are flawed both on the observational and the theoretical side.

\end{itemize}

In an upcoming paper (Rossetti \& Molendi 2009) we will investigate further the issue of medium and high  entropy core
systems with a large central abundance; we will do this by performing bi-dimensional analysis of a smaller sample of bright and nearby clusters.


\begin{acknowledgements}

This research has made use of two databases: the NASA/IPAC Extragalactic Database (NED) and
the X-Rays Clusters Database (BAX) and of three archives:
the High Energy Astrophysics Science Archive Research Center (HEASARC); the \emph{XMM-Newton} Science Archive (XSA) and
the Archive of \emph{Chandra} Cluster Entropy Profile Tables (ACCEPT).
We would like to express our appreciation for the excellent work by
Cavagnolo and collaborators in setting up the ACCEPT archive.
We acknowledge useful discussions with Stefano Ettori, Stefano Borgani.
We thank Sabrina De Grandi and Fabio Gastaldello for a careful and critical reading of
the manuscript.

\end{acknowledgements}

\bibliographystyle{aa}
\bibliography{paper4}
\begin{appendix} \label{app: A}
\section{}
In this Appendix we compare $E\!M_\mathrm{OUT}$, the emission measure from the 0.05-0.20~$R_{180}$ 
ring, with another proxy for the emission measure at large radii, namely 
the self-similar scaling factor $E\!M_\mathrm{SSS}$ which is defined as:

\begin{equation}
E\!M_\mathrm{SSS} \equiv \Delta_\mathrm{z}^{3/2} \, (1+z)^{9/2} \, (\mathrm{k}T_\mathrm{OUT}/10 \,
\mathrm{keV})^{1/2} ,
\label{eq: EM_SSS}
\end{equation}
for details about this definition we refer our readers to \citet{arnaud02b}.
As a reference value of the  emission measure at large radii we employ 
$E\!M_0$, the emission measure calculated in the 0.2-0.4~$R_{180}$ ring.
$E\!M_0$ is available for the subsample of distant ($z > 0.1$) clusters, for which it is
possible to measure $E\!M$ out to 0.4~$R_{180}$
(the values of $E\!M_0$  are taken from \citealt{leccardi08b}).
Since we are interested in emission measure ratios, 
for each distant cluster, we calculated our ``ideal'' ratio $E\!M_\mathrm{IN}/E\!M_0$,
determined directly from the data, the self-similar scaled ratio
$E\!M_\mathrm{IN}/E\!M_\mathrm{SSS}$, and the standard ratio
$E\!M_\mathrm{IN}/E\!M_\mathrm{OUT}$.
\begin{figure}
  \centering
  \begin{tabular}{cc}
     \hspace{-4mm}\resizebox{4.8cm}{!}{\includegraphics{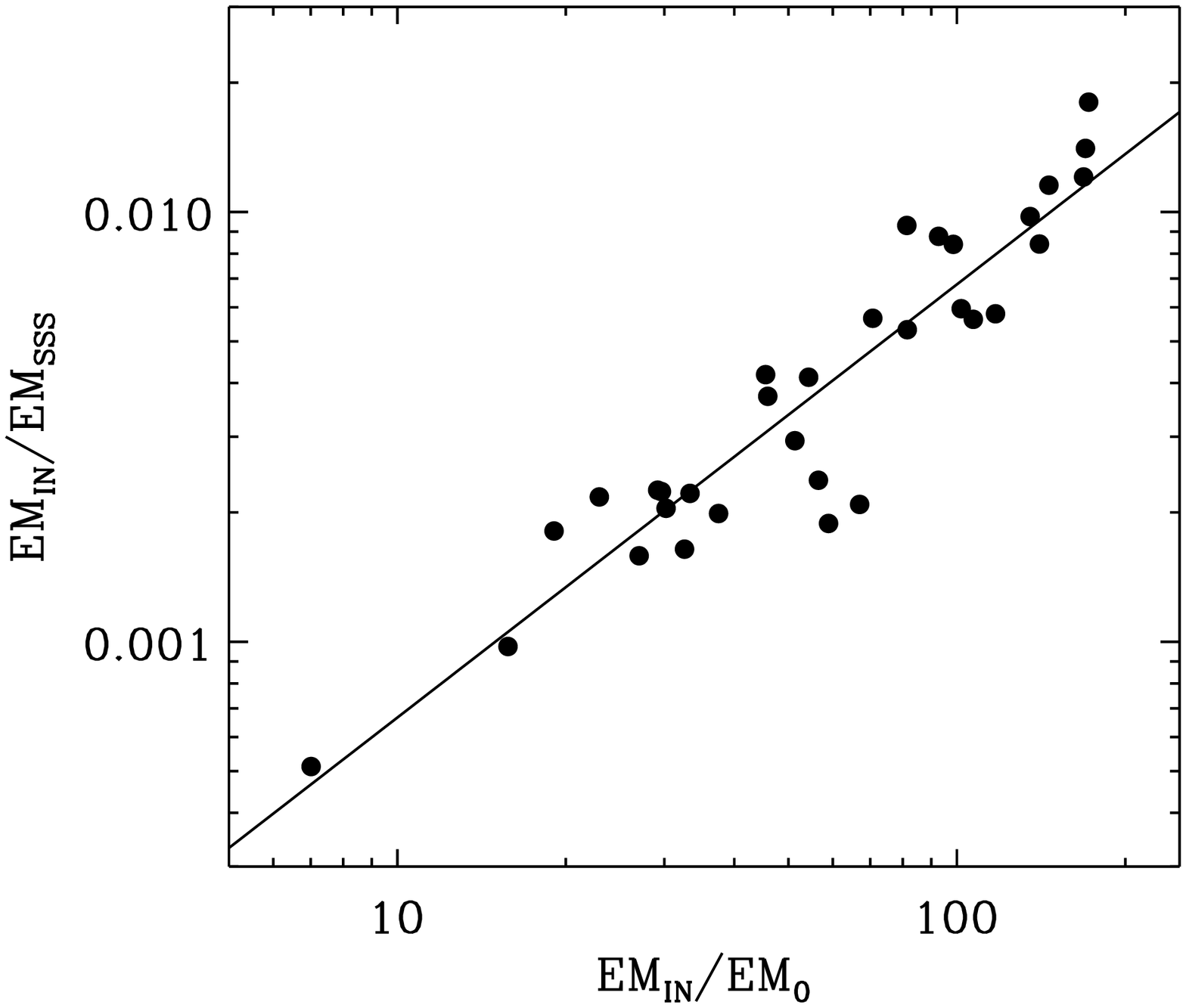}} &
     \hspace{-8mm}\resizebox{4.8cm}{!}{\includegraphics{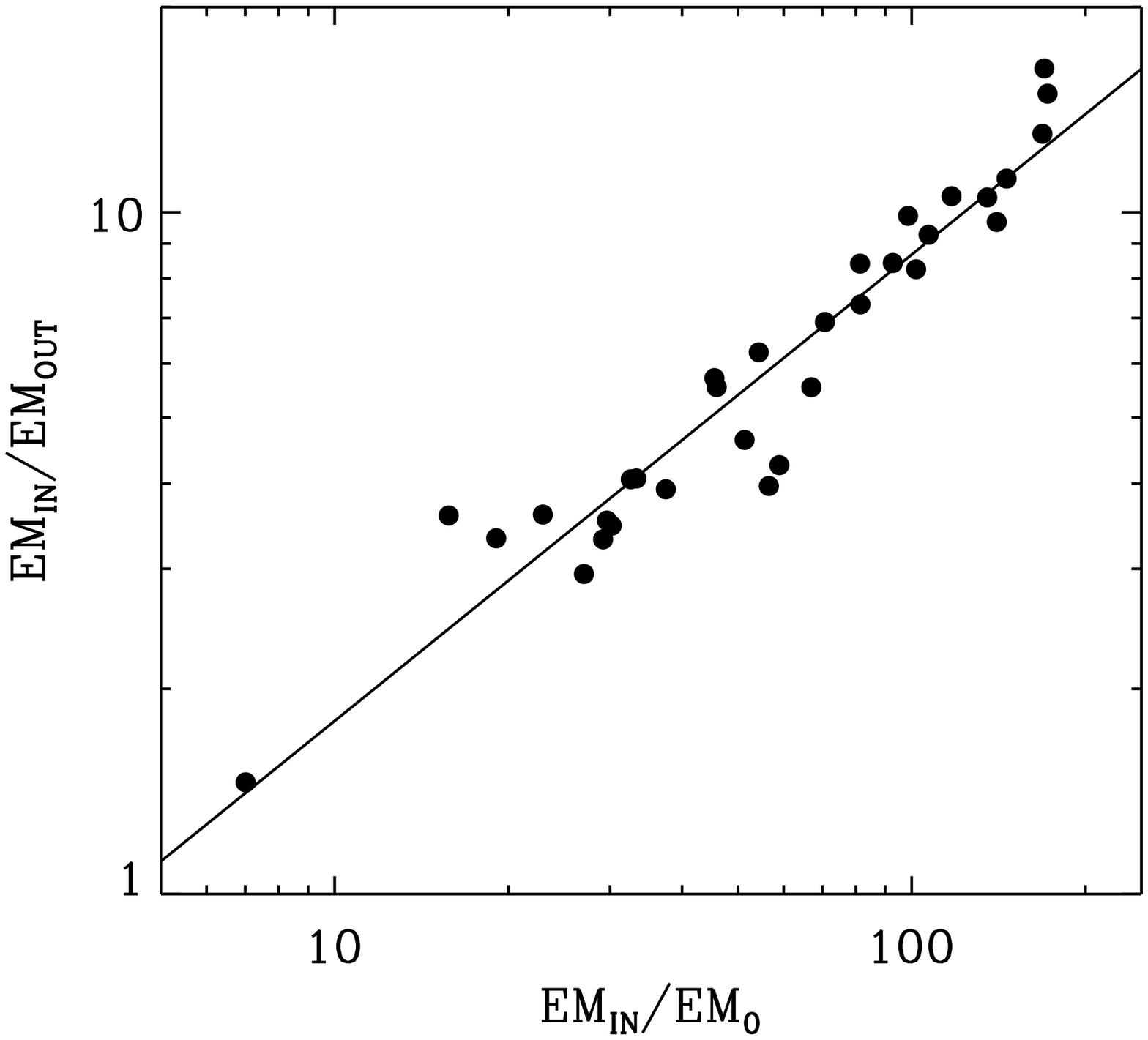}} \\
 \end{tabular}
  \caption{Correlations between various definitions for the $E\!M$ ratios (see text for
  details).
  Left panel: self-similar scaled vs. ideal ratios. Right panel: standard vs. ideal ratios.
  The solid curve is the best fit with a power law. The uncertainties are smaller than the
  point size. The scatter is 29\% and 16\% for left and right panels respectively.}
  \label{fig: ratio_SB}
\end{figure}
In Fig.~\ref{fig: ratio_SB} we compare the self-similar scaled (left panel) and the standard
(right panel) ratios to the ideal ratio.
For both cases we find a good correlation, but the scatter is smaller (16\% vs. 29\%) when
using the standard ratio.
Throughout this paper we make use of $E\!M_\mathrm{IN}/E\!M_\mathrm{OUT}$,
but we emphasize that our results are largely independent of this particular choice.
Indeed, the appearance of the plot in Fig.~\ref{fig: SB_T plot} is very similar  when using
\begin{equation}
\tilde\sigma \equiv (T_\mathrm{IN}/T_\mathrm{OUT}) \times (E\!M_\mathrm{IN}/E\!M_\mathrm{SSS})^{-1/3} ,
\end{equation}
which differs from $\sigma$ for the use of the self-similar scaling.
\begin{figure}
  \centering
  \hspace{-0.82 truecm}\resizebox{9.8 truecm}{!}{\includegraphics{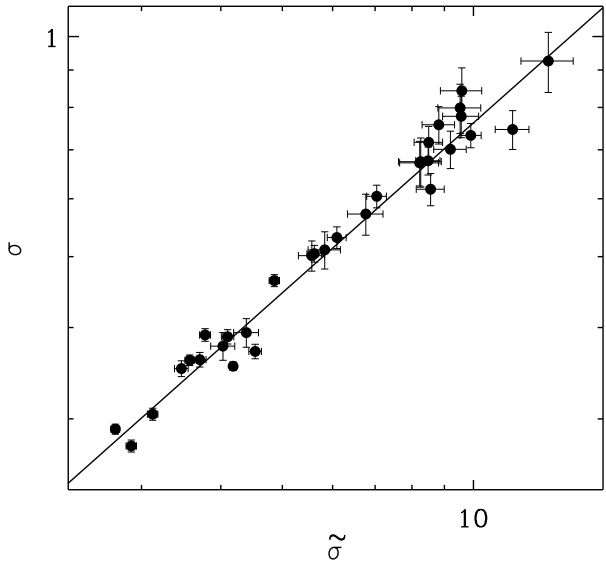}} \\
  \caption{Correlation between $\sigma$ and $\tilde\sigma$ (see text for definitions) for
  clusters with $z > 0.1$. The solid line shows the best-fit power law; the scatter is
  $\approx$~6\%.
}
  \label{fig: entropy ratio}
\end{figure}
In Fig.~\ref{fig: entropy ratio} we show the correlation between $\sigma$ and $\tilde\sigma$
for the subsample of distant clusters; we find a small, $\approx$~6\%, scatter around the best-fit
power law (represented with a solid line); this is another confirmation of the robustness of
our entropy indicator.

\end{appendix}

%
\longtab{1}{
\centering
\begin{longtable}{l r r r r l}
  \caption{\label{tab: sample} Physical properties and observation details for the 59 clusters
  of our sample.}\\
    \hline \hline
    Name & $z\mathrm{^a}$ & $kT_\mathrm{OUT}\mathrm{^b}$
    & Exp. time$\mathrm{^c}$ & $R_\mathrm{SB}\mathrm{^d}$ & Filter \\
    \hline
    Abell 4038  	  & 0.0300 &  3.0 &  78.1 & 1.45 & MEDIUM \\
    Abell 2199  	  & 0.0301 &  4.1 &  38.4 & 1.14 & THIN1 \\
    2A 0335+096		  & 0.0349 &  3.6 & 230.8 & 1.02 & THIN1 \\
    Abell 2052  	  & 0.0355 &  2.9 &  85.1 & 0.99 & THIN1 \\
    Abell 576   	  & 0.0390 &  3.8 &  42.7 & 1.54 & MEDIUM \\
    Abell 3571  	  & 0.0391 &  6.3 &  43.5 & 1.54 & MEDIUM \\
    Abell 119   	  & 0.0442 &  6.0 &  54.1 & 1.62 & THIN1 \\
    MKW 03		  & 0.0450 &  3.3 &  99.2 & 1.08 & THIN1 \\
    Abell 3376            & 0.0456 &  3.9 &  56.4 & 1.10 & MEDIUM \\
    Abell 1644  	  & 0.0470 &  4.2 &  42.0 & 1.40 & THIN1 \\
    Abell 4059  	  & 0.0475 &  4.0 &  64.9 & 1.11 & THIN1 \\
    Abell 3558  	  & 0.0480 &  5.2 & 126.4 & 1.00 & THICK \\
    Abell 3562  	  & 0.0480 &  4.3 & 116.6 & 1.85 & THIN1 \\
    Triangulum Austr.	  & 0.0510 &  9.2 &  27.3 & 1.15 & MEDIUM \\
    Hydra A		  & 0.0538 &  3.4 &  52.2 & 1.59 & THIN1 \\
    Abell 754             & 0.0542 &  8.7 &  30.8 & 1.11 & MEDIUM\\
    Abell 85    	  & 0.0551 &  5.5 &  34.6 & 1.04 & MEDIUM \\
    Abell 2319  	  & 0.0557 &  9.2 &  44.6 & 1.80 & MEDIUM \\
    Abell 3158  	  & 0.0597 &  4.9 &  54.0 & 1.58 & THIN1 \\
    Abell 1795  	  & 0.0625 &  5.4 &  97.3 & 1.29 & THIN1 \\
    Abell 399   	  & 0.0720 &  6.0 &  27.9 & 1.84 & THIN1 \\
    Abell 401   	  & 0.0740 &  7.3 &  34.7 & 1.76 & MEDIUM \\
    Abell 3112  	  & 0.0750 &  4.3 &  64.6 & 1.23 & MEDIUM \\
    Abell 2029  	  & 0.0773 &  6.2 &  30.8 & 1.17 & THIN1 \\
    Abell 2255  	  & 0.0806 &  6.2 &  25.1 & 1.37 & THIN1 \\
    Abell 1650  	  & 0.0838 &  5.4 &  75.0 & 1.29 & MEDIUM \\
    Abell 2597  	  & 0.0852 &  3.5 & 144.3 & 1.07 & THIN1 \\
    Abell S0084 	  & 0.1080 &  3.3 &  44.8 & 1.28 & THIN1 \\
    Abell 2034  	  & 0.1130 &  7.0 &  27.9 & 1.16 & THIN1 \\
    Abell 2051	          & 0.1150 &  3.8 &  83.6 & 1.08 & THIN1 \\
    Abell 3814	          & 0.1179 &  3.3 &  71.2 & 1.11 & THIN1 \\
    Abell 2050	          & 0.1183 &  5.3 &  77.3 & 1.13 & THIN1 \\
    RXCJ1141.4-1216	  & 0.1195 &  3.8 &  82.0 & 1.03 & THIN1 \\
    Abell 1084	          & 0.1323 &  3.9 &  72.4 & 1.03 & THIN1 \\
    Abell 1068  	  & 0.1375 &  4.5 &  56.3 & 1.09 & MEDIUM \\
    Abell 3856	          & 0.1379 &  6.4 &  54.6 & 1.11 & THIN1 \\
    Abell 3378	          & 0.1410 &  4.9 &  58.3 & 1.07 & THIN1 \\
    Abell 0022	          & 0.1424 &  5.7 &  41.9 & 1.02 & THIN1 \\
    Abell 1413  	  & 0.1427 &  6.7 &  71.7 & 1.10 & THIN1 \\
    Abell 2328	          & 0.1470 &  5.6 &  71.1 & 1.07 & THIN1 \\
    Abell 3364	          & 0.1483 &  6.7 &  67.1 & 1.12 & THIN1 \\
    Abell 2204  	  & 0.1522 &  8.5 &  51.2 & 1.06 & MEDIUM \\
    Abell 0907	          & 0.1527 &  6.1 &  22.9 & 1.16 & THIN1 \\
    Abell 3888	          & 0.1529 &  8.6 &  42.8 & 1.31 & THIN1 \\
    RXCJ2014.8-2430	  & 0.1612 &  7.1 &  64.8 & 1.05 & THIN1 \\
    Abell 3404	          & 0.1670 &  7.1 &  59.2 & 1.11 & THIN1 \\
    Abell 1914  	  & 0.1712 &  8.7 &  62.9 & 1.17 & THIN1 \\
    Abell 2218  	  & 0.1756 &  6.5 & 117.0 & 1.17 & THIN1 \\
    Abell 1689  	  & 0.1832 &  9.2 & 106.7 & 1.14 & THIN1 \\
    Abell 383   	  & 0.1871 &  4.4 &  82.3 & 1.33 & MEDIUM \\
    Abell 115N   	  & 0.1971 &  5.1 & 103.2 & 1.20 & MEDIUM \\
    Abell 2163  	  & 0.2030 & 15.5 &  29.2 & 1.07 & THIN1 \\
    Abell 963   	  & 0.2060 &  6.5 &  69.4 & 1.19 & MEDIUM \\
    Abell 209   	  & 0.2060 &  6.6 &  49.3 & 1.19 & MEDIUM \\
    Abell 773   	  & 0.2170 &  7.5 &  45.6 & 1.16 & MEDIUM \\
    Abell 1763  	  & 0.2230 &  7.2 &  36.3 & 1.08 & MEDIUM \\
    Abell 2390  	  & 0.2280 & 11.2 &  29.4 & 1.11 & THIN1 \\
    Abell 2667  	  & 0.2300 &  7.7 &  59.9 & 1.48 & MEDIUM \\
    RX J2129.3+0005	  & 0.2350 &  5.5 & 102.0 & 1.21 & MEDIUM \\
    \hline
  \end{longtable}
  \begin{list}{}{}
    \item[Notes:] $\mathrm{^a}$~redshift taken from the NASA Extragalactic
    Database; $\mathrm{^b}$~reference temperature in keV derived from our
    analysis (see~Sect.~\ref{sec: sp ana}); $\mathrm{^c}$~total good exposure
    time in ks; $\mathrm{^d}$~intensity of residual soft protons
    (see~Sect.~\ref{sec: spec prep}).
    \end{list}
}

\longtab{2}{
\centering
\begin{longtable}{l c c c }
  \caption{\label{tab: sigma_z} Pseudo-entropy ratio and metal abundances for the inner region of the 59 objects in our sample.}\\
    \hline \hline
    Name & $\sigma\mathrm{^a}$ & Entropy               & $Z_\mathrm{IN}/Z_\odot\mathrm{^c}$\\
         &                     & Classif.$\mathrm{^b}$ &                                   \\
    \hline
    Abell 4038   	    &  $ 0.499 \pm 0.004 $  &  MEC &  $ 0.41 \pm 0.01 $  \\
    Abell 2199  	    &  $ 0.414 \pm 0.003 $  &  LEC &  $ 0.45 \pm 0.01 $  \\
    2A 0335+096		    &  $ 0.262 \pm 0.001 $  &  LEC &  $ 0.66 \pm 0.00 $  \\
    Abell 2052  	    &  $ 0.420 \pm 0.003 $  &  LEC &  $ 0.55 \pm 0.01 $  \\
    Abell 576   	    &  $ 0.609 \pm 0.013 $  &  MEC &  $ 0.53 \pm 0.04 $  \\
    Abell 3571  	    &  $ 0.608 \pm 0.008 $  &  MEC &  $ 0.37 \pm 0.02 $  \\
    Abell 119   	    &  $ 0.734 \pm 0.024 $  &  HEC &  $ 0.28 \pm 0.05 $  \\
    MKW 03		    &  $ 0.477 \pm 0.004 $  &  MEC &  $ 0.52 \pm 0.01 $  \\
    Abell 3376              &  $ 0.682 \pm 0.025 $  &  HEC &  $ 0.30 \pm 0.05 $  \\
    Abell 1644  	    &  $ 0.519 \pm 0.013 $  &  MEC &  $ 0.74 \pm 0.05 $  \\
    Abell 4059  	    &  $ 0.434 \pm 0.004 $  &  LEC &  $ 0.60 \pm 0.02 $  \\
    Abell 3558  	    &  $ 0.605 \pm 0.006 $  &  MEC &  $ 0.38 \pm 0.01 $  \\
    Abell 3562  	    &  $ 0.601 \pm 0.008 $  &  MEC &  $ 0.50 \pm 0.02 $  \\
    Triangulum Austr.       &  $ 0.675 \pm 0.016 $  &  HEC &  $ 0.29 \pm 0.03 $  \\
    Hydra A		    &  $ 0.406 \pm 0.004 $  &  LEC &  $ 0.34 \pm 0.01 $  \\
    Abell 754               &  $ 0.514 \pm 0.011 $  &  MEC &  $ 0.31 \pm 0.03 $  \\
    Abell 85    	    &  $ 0.379 \pm 0.005 $  &  LEC &  $ 0.51 \pm 0.02 $  \\
    Abell 2319  	    &  $ 0.572 \pm 0.010 $  &  MEC &  $ 0.26 \pm 0.02 $  \\
    Abell 3158  	    &  $ 0.733 \pm 0.015 $  &  HEC &  $ 0.32 \pm 0.03 $  \\
    Abell 1795  	    &  $ 0.345 \pm 0.002 $  &  LEC &  $ 0.40 \pm 0.01 $  \\
    Abell 399   	    &  $ 0.694 \pm 0.031 $  &  HEC &  $ 0.27 \pm 0.06 $  \\
    Abell 401   	    &  $ 0.692 \pm 0.021 $  &  HEC &  $ 0.25 \pm 0.03 $  \\
    Abell 3112  	    &  $ 0.376 \pm 0.004 $  &  LEC &  $ 0.59 \pm 0.02 $  \\
    Abell 2029  	    &  $ 0.441 \pm 0.006 $  &  LEC &  $ 0.47 \pm 0.02 $  \\
    Abell 2255  	    &  $ 0.865 \pm 0.069 $  &  HEC &  $ 0.17 \pm 0.12 $  \\
    Abell 1650  	    &  $ 0.516 \pm 0.009 $  &  MEC &  $ 0.40 \pm 0.02 $  \\
    Abell 2597  	    &  $ 0.346 \pm 0.002 $  &  LEC &  $ 0.36 \pm 0.01 $  \\
    Abell S0084  	    &  $ 0.618 \pm 0.031 $  &  MEC &  $ 0.71 \pm 0.10 $  \\
    Abell 2034  	    &  $ 0.778 \pm 0.051 $  &  HEC &  $ 0.47 \pm 0.11 $  \\
    Abell 2051	            &  $ 0.746 \pm 0.046 $  &  HEC &  $ 0.27 \pm 0.10 $  \\
    Abell 3814	            &  $ 0.371 \pm 0.008 $  &  LEC &  $ 0.66 \pm 0.04 $  \\
    Abell 2050	            &  $ 0.733 \pm 0.028 $  &  HEC &  $ 0.32 \pm 0.05 $  \\
    RXCJ1141.4-1216	    &  $ 0.354 \pm 0.005 $  &  LEC &  $ 0.64 \pm 0.03 $  \\
    Abell 1084	            &  $ 0.361 \pm 0.006 $  &  LEC &  $ 0.42 \pm 0.02 $  \\
    Abell 1068  	    &  $ 0.304 \pm 0.006 $  &  LEC &  $ 0.48 \pm 0.02 $  \\
    Abell 3856              &  $ 0.605 \pm 0.022 $  &  MEC &  $ 0.34 \pm 0.05 $  \\
    Abell 3378	            &  $ 0.389 \pm 0.009 $  &  LEC &  $ 0.56 \pm 0.03 $  \\
    Abell 0022	            &  $ 0.701 \pm 0.041 $  &  HEC &  $ 0.31 \pm 0.08 $  \\
    Abell 1413  	    &  $ 0.505 \pm 0.013 $  &  MEC &  $ 0.33 \pm 0.03 $  \\
    Abell 2328	            &  $ 0.926 \pm 0.087 $  &  HEC &  $ 0.35 \pm 0.18 $  \\
    Abell 3364	            &  $ 0.676 \pm 0.030 $  &  HEC &  $ 0.35 \pm 0.06 $  \\
    Abell 2204  	    &  $ 0.290 \pm 0.005 $  &  LEC &  $ 0.51 \pm 0.02 $  \\
    Abell 0907	            &  $ 0.394 \pm 0.018 $  &  LEC &  $ 0.46 \pm 0.07 $  \\
    Abell 3888	            &  $ 0.717 \pm 0.037 $  &  HEC &  $ 0.26 \pm 0.07 $  \\
    RXCJ2014.8-2430	    &  $ 0.275 \pm 0.005 $  &  LEC &  $ 0.45 \pm 0.02 $  \\
    Abell 3404	            &  $ 0.507 \pm 0.018 $  &  MEC &  $ 0.30 \pm 0.05 $  \\
    Abell 1914  	    &  $ 0.577 \pm 0.018 $  &  MEC &  $ 0.29 \pm 0.04 $  \\
    Abell 2218  	    &  $ 0.791 \pm 0.031 $  &  HEC &  $ 0.28 \pm 0.06 $  \\
    Abell 1689  	    &  $ 0.464 \pm 0.009 $  &  MEC &  $ 0.31 \pm 0.02 $  \\
    Abell 383   	    &  $ 0.361 \pm 0.008 $  &  LEC &  $ 0.52 \pm 0.03 $  \\
    Abell 115N   	    &  $ 0.329 \pm 0.009 $  &  LEC &  $ 0.42 \pm 0.03 $  \\
    Abell 2163  	    &  $ 0.720 \pm 0.046 $  &  HEC &  $ 0.34 \pm 0.10 $  \\
    Abell 963   	    &  $ 0.572 \pm 0.037 $  &  MEC &  $ 0.51 \pm 0.10 $  \\
    Abell 209   	    &  $ 0.531 \pm 0.018 $  &  MEC &  $ 0.36 \pm 0.05 $  \\
    Abell 773   	    &  $ 0.672 \pm 0.047 $  &  HEC &  $ 0.30 \pm 0.10 $  \\
    Abell 1763  	    &  $ 0.674 \pm 0.052 $  &  HEC &  $ 0.52 \pm 0.15 $  \\
    Abell 2390  	    &  $ 0.377 \pm 0.016 $  &  LEC &  $ 0.42 \pm 0.05 $  \\
    Abell 2667  	    &  $ 0.351 \pm 0.009 $  &  LEC &  $ 0.47 \pm 0.03 $  \\
    RX~J2129.3+0005	    &  $ 0.391 \pm 0.008 $  &  LEC &  $ 0.35 \pm 0.03 $  \\
    \hline
  \end{longtable}
  \begin{list}{}{}
    \item[Notes:] $\mathrm{^a}$~pseudo-entropy ratio, see Sect.~\ref{sec: entropy} for details;
                  $\mathrm{^b}$~pseudo entropy class, see Sect.~\ref{sec: compare core} for details;
                  $\mathrm{^c}$~metal abundance in central region, see Sect.~\ref{sec: Z vs S} for details.
    \end{list}
}

\longtab{3}{
\centering
\begin{longtable}{l c c c c c c r r r}
  \caption{\label{tab: classif} Classification of objects based on radio, optical and X-ray properties.}\\
    \hline \hline
    Name & R$\mathrm{^a}$ & R.ref & O$\mathrm{^b}$& O.ref & X-ray$\mathrm{^c}$ & X-ray ref & Note & Core Cl. & Dyn.Cl.\\
    \hline
    Abell 4038   	  & H & Sl98,Sl01   &  ? & Bu04      & NCC        & Sn06           & Y & INT & NOM \\ 
    Abell 2199  	  & - &             &  ? & Oe01      & CC         & Jo02           & N & CC  & NOM \\ 
    2A 0335+096		  & - &             &  - &           & CC         & Ma03           & N & CC  & NOM \\ 
    Abell 2052  	  & - &             &  N & Di87,Ma92 & CC         & Bl03           & N & CC  & NOM \\ 
    Abell 576   	  & N &  Fe07       &  - &           & MRG \& NCC & Mo96,Ke04,Du07 & N & NCC & MRG \\ 
    Abell 3571  	  & N &  Ve02       &  - &           & NCC        & Oh06,Sn06,Da08 & Y & INT & NOM \\ 
    Abell 119   	  & N &  Mu04       &  - &           & MRG \& NCC & Sr06,Ro06      & N & NCC & MRG \\ 
    MKW 03		  & - &             &  - &           & CC         & Ma02,Su07      & N & CC  & NOM \\ 
    Abell 3376            & R &  Ba06       &  N & Gi97      & MRG \& NCC & Ro06,Ca09      & N & NCC & MRG \\ 
    Abell 1644  	  & - &             &  N & Tu01      & INT        & Re04,Du05      & N & INT & NOM \\ 
    Abell 4059  	  & - &             &  - &           & CC         & Ch04,Ca09      & N & CC  & NOM \\ 
    Abell 3558  	  & N &  Ba98       &  Y & Ba98      & INT        & Ro07           & N & INT & NOM \\ 
    Abell 3562  	  & H &  Gi05       &  Y & Ba98      & MRG \& NCC & Fi04           & N & NCC & MRG \\ 
    Triangulum Austr.     & - &             &  - &           & NCC        & Ch07           & N & NCC & NOM \\ 
    Hydra A		  & - &             &  - &           & CC         & Da01           & N & CC  & NOM \\ 
    Abell 754             & H & Ka01,Ba03   &  ? & Ro98      & MRG \& NCC & Kr03,He04      & N & NCC & MRG \\ 
    Abell 85    	  & - &             &  N & Ma92,Br09 & MRG \& CC  & Ke02,Dr05      & N & CC  & MRG \\ 
    Abell 2319  	  & H & Fe97,Ca06   &  Y & Oe95      & MRG \& NCC & Oh04           & N & NCC & MRG \\ 
    Abell 3158  	  & - &             &  Y & Jo08      & MRG \& NCC & Sn08,Gh09      & N & NCC & MRG \\ 
    Abell 1795  	  & - &             &  - &           & CC         & Et02           & N & CC  & NOM \\ 
    Abell 399   	  & N &   Gi06      &  Y & Gi97      & MRG \& NCC & Sa04,Ro06      & N & NCC & MRG \\ 
    Abell 401   	  & H &   Ba03      &  Y & Gi97      & MRG \& NCC & Sa04,Bo08      & N & NCC & MRG \\ 
    Abell 3112  	  & - &             &  - &           & CC         & Ta03           & N & CC  & NOM \\ 
    Abell 2029  	  & - &             &  - &           & CC         & Cl04           & N & CC  & NOM \\ 
    Abell 2255  	  & H &  Pi08       &  - &           & MRG \& NCC & Sa06           & N & NCC & MRG \\ 
    Abell 1650  	  & - &             &  - &           & CC         & Do05,Ca09      & Y & INT & NOM \\ 
    Abell 2597  	  & - &             &  - &           & CC         & Mo05,Ca09      & N & CC  & NOM \\ 
    Abell S0084  	  & - &             &  - &           & INT        & Si09           & N & INT & NOM \\ 
    Abell 2034  	  & - &             &  - &           & MRG \& NCC & Ke03,Bl07      & N & NCC & MRG \\ 
    Abell 2051	          & - &             &  - &           & NCC        & Pr07,Cr08      & N & NCC & NOM \\ 
    Abell 3814	          & - &             &  - &           & CC         & Cr08,Le08      & N & CC  & NOM \\ 
    Abell 2050	          & - &             &  - &           & NCC        & Pr07,Cr08      & N & NCC & NOM \\ 
    RXCJ1141.4-1216	  & - &             &  - &           & CC         & Pr07,Cr08      & N & CC  & NOM \\ 
    Abell 1084	          & - &             &  - &           & CC         & Pr07,Cr08      & N & CC  & NOM \\ 
    Abell 1068  	  & - &             &  - &           & CC         & Wi04           & N & CC  & NOM \\ 
    Abell 3856            & - &             &  - &           & NCC        & Pr07,Cr08      & N & NCC & NOM \\ 
    Abell 3378	          & - &             &  - &           & CC         & Pr07,Cr08      & N & CC  & NOM \\ 
    Abell 0022	          & - &             &  - &           & NCC        & Pr07,Cr08      & N & NCC & NOM \\ 
    Abell 1413  	  & - &             &  - &           & INT        & Vi05,Bl07,Ca09 & N & INT & NOM \\ 
    Abell 2328	          & - &             &  - &           & NCC        & Pr07,Cr08      & N & NCC & NOM \\ 
    Abell 3364	          & - &             &  - &           & NCC        & Pr07,Cr08      & N & NCC & NOM \\ 
    Abell 2204  	  & - &             &  - &           & CC         & Sa09           & N & CC  & NOM \\ 
    Abell 0907	          & - &             &  - &           & CC         & Vi05,Cr08      & N & CC  & NOM \\ 
    Abell 3888	          & - &             &  - &           & NCC        & Cr08           & N & NCC & NOM \\ 
    RXCJ2014.8-2430	  & - &             &  - &           & CC         & Cr08           & N & CC  & NOM \\ 
    Abell 3404	          & - &             &  - &           & INT        & Cr08           & N & INT & NOM \\ 
    Abell 1914  	  & H & Ba03        &  - &           & MRG \& NCC & Go04,Bl07      & N & NCC & MRG \\ 
    Abell 2218  	  & H & Gi00        &  Y &  Gi97     & MRG \& NCC & Go04,Bl07      & N & NCC & MRG \\ 
    Abell 1689  	  & - &             &  ? & Gi01,Lo06 & MRG \& CC  & Pe98,An04,Ca09 & Y & INT & NOM \\ 
    Abell 383   	  & - &             &  - &           & CC         & Vi05,Ca09      & N & CC  & NOM \\ 
    Abell 115N   	  & ? &  Go01       &  Y & Ba07b     & MRG \& CC  & Gu05           & N & CC  & MRG \\ 
    Abell 2163  	  & H &  Fe01       &  Y & Ma08      & MRG \& NCC & Go04           & N & NCC & MRG \\ 
    Abell 963   	  & N &  Ca08       &  - &           & NCC        & Sm05,Bl07,Ca09 & Y & INT & NOM \\ 
    Abell 209   	  & H &  Ve07       &  Y & Da02,Me04 & NCC        & Ca09           & N & NCC & MRG \\ 
    Abell 773   	  & H &  Go01       &  Y & Ba07a     & MRG \& NCC & Go04,Ca09      & N & NCC & MRG \\ 
    Abell 1763  	  & N &  Ve08       &  Y & Fa08      & MRG \& NCC & Du08,Ca09      & N & NCC & MRG \\ 
    Abell 2390  	  & M &  Ba03       &  N & Le91      & CC         & Vi05,Ca09      & N & CC  & NOM \\ 
    Abell 2667  	  & N &  Ve08       &  - &           & CC         & Co06,Ca09      & N & CC  & NOM \\ 
    RX~J2129.3+0005	  & - &             &  - &           & CC         & Ca09           & N & CC  & NOM \\ 
    \hline
  \end{longtable}

  \begin{list}{}{}
    \item[Notes:] $\mathrm{^a}$~radio emission; $\mathrm{^b}$~optical substructure;
    $\mathrm{^c}$~X-ray classification; a detailed description
    of the table is provided in Sect.~\ref{sec: class sch}.
    \item[References] divided by band of electromagnetic spectrum.\\
  {\bf References for radio observations}: (Sl98) \citealt{Sl98}; (Sl01) \citealt{Sl01}; (Fe07) \citealt{Fe07};
  (Ve02) \citealt{Ve02}; (Mu04) \citealt{Mu04}; (Ba06) \citealt{Ba06}; (Ba98) \citealt{Ba98};
  (Gi05) \citealt{Gi05}; (Ka01) \citealt{Ka01}; (Ba03) \citealt{Ba03}; (Fe97) \citealt{Fe97};
  (Ca06) \citealt{Ca06}; (Gi06) \citealt{Gi06}; (Pi08) \citealt{Pi08}; (Gi00) \citealt{Gi00};
  (Go01) \citealt{Go01}; (Fe01) \citealt{Fe01}; (Ca08) \citealt{Ca08}; (Ve07) \citealt{Ve07};
  (Ve08) \citealt{Ve08}.\\
  {\bf References for optical observations}: (Bu04) \citealt{Bu04}; (Oe01) \citealt{Oe01}; (Di87) \citealt{Di87};
  (Ma92) \citealt{Ma92}; (Br09) \citealt{Br09}; (Gi97) \citealt{Gi97}; (Tu01) \citealt{Tu01}; (Ro98) \citealt{Ro98};
  (Oe95) \citealt{Oe95};
  (Jo08) \citealt{Jo08}; (Gi01) \citealt{Gi01}; (Lo06) \citealt{Lo06}; (Ba07b) \citealt{Ba07b}; (Ma08) \citealt{Ma08};
  (Ba07a) \citealt{Ba07a}; (Fa08) \citealt{Fa08}; (Le91) \citealt{Le91}.\\
  {\bf References for X-ray observations}: (Sa06) \citealt{Sa06}; (Jo02) \citealt{Jo02}; (Ma03) \citealt{Ma03};
  (Bl03) \citealt{Bl03}; (Mo96) \citealt{Mo96}; (Ke04) \citealt{Ke04}; (Du07) \citealt{Du07}; (Oh06) \citealt{Oh06};
  (Sn06) \citealt{Sn06}; (Da08) \citealt{Da08}; (Sr06) \citealt{Sr06}; (Ro06) \citealt{Ro06};
  (Ma02) \citealt{mazzotta02}; (Su07) \citealt{Su07}; (Re04) \citealt{Re04}; (Du05) \citealt{Du05};
  (Ch04) \citealt{Ch04};  (Ro07) \citealt{rossetti07}; (Fi04) \citealt{Fi04}; (Ch07) \citealt{Ch07};
  (Da01) \citealt{Da01}; (Kr03) \citealt{Kr03}; (He04) \citealt{He04}; (Ke02) \citealt{Ke02}; (Dr05) \citealt{Dr05};
  (Oh04) \citealt{Oh04}; (Sn08) \citealt{snowden08}; (Gh09) \citealt{Gh09}; (Et02) \citealt{Et02};
  (Sa04) \citealt{Sa04}; (Bo08) \citealt{bourdin08};
  (Ta03) \citealt{Ta03}; (Cl04) \citealt{Cl04};
  (Sa06) \citealt{Sa06}; (Do05) \citealt{Do05}; (Mo05) \citealt{Mo05}; (Si09) \citealt{Si09};
  (Ke03) \citealt{Ke03}; (Cr08) \citealt{Cr08}; (Le08) \citealt{leccardi08b}; (Pr07) \citealt{pratt07};
  (Wi04) \citealt{Wi04}; (Vi05) \citealt{vikh05}; (Bl07) \citealt{baldi07}; (Go04) \citealt{Go04};
  (Pe98) \citealt{peres98}; (An04) \citealt{An04}; (Gu05) \citealt{Gu05}; (Sm05) \citealt{Sm05};
  (Da02) \citealt{Da02}; (Me04) \citealt{Me04}; (Du08) \citealt{Du08}; (Co06) \citealt{Co06};
  (Ca09) \citealt{Ca09}.
    \end{list}

}

\end{document}